\documentclass[journal]{IEEEtran}
\usepackage{amsmath}
\usepackage{graphicx}
\usepackage{epstopdf}
\usepackage{latexsym}
\usepackage{amssymb}
\usepackage{cite}
\usepackage{ushort}
\usepackage{mathtools}
\usepackage{caption}
\usepackage{url}
\usepackage{pgfplots}
\usepackage{tikzscale}
\usepackage{booktabs}
\usepackage{ushort}
\usepackage{calrsfs}
\usepackage{makecell}
\usepackage{array}
\usepackage{textcomp}
\usepackage{calc}
\usepackage{tikz}
\usepackage{cases}
\usepackage{empheq}
\usetikzlibrary{shapes,arrows}
\usetikzlibrary{calc}
\usetikzlibrary{positioning}
\usetikzlibrary{arrows}
\usetikzlibrary{arrows.meta}
\usetikzlibrary{arrows,scopes}
\usetikzlibrary{backgrounds}
\usepackage{xfrac} 
\usepackage{nicefrac}  
\usepackage{upgreek}
\usepackage{xcolor}
\usepackage{siunitx}

\def\a{\mathbf{a}}

\def\x{\mathbf{x}}
\def\y{\mathbf{y}}
\def\v{\mathbf{v}}
\def\A{\mathbf{A}}

\def\w{\mathbf{w}}
\def\h{\mathbf{h}}
\def\H{\mathbf{H}}
\def\g{\mathbf{g}}
\def\B{\mathbf{B}}

\def\u{{\mathbf{u}}}

\def\0{{\mathbf{0}}}
\def\A{\mathbf{A}}

\def\bPsi{\boldsymbol{\Psi}}
\def\bpsi{\boldsymbol{\psi}}

\def\bOmega{\boldsymbol{\Omega}}
\def\bphi{\boldsymbol{\upvarphi}}

\def\k{\mathbf{k}}
\def\I{\mathbf{I}}

\DeclareMathOperator {\Diag}{Diag}

\DeclareMathOperator {\tr}{tr}

\DeclareMathOperator {\Exp}{E}

\DeclareMathAlphabet{\pazocal}{OMS}{zplm}{m}{n}

\newcommand{\+}{\hskip 0.13em {+}\hskip 0.13em}
\renewcommand{\-}{\hskip 0.13em{-}\hskip 0.13em}

\newcommand*{\transp}{{\scriptscriptstyle{T}}}
\newcommand*{\herm}{{\scriptscriptstyle{H}}}
\newcommand*{\inv}{{\scriptscriptstyle{-1}}}

\tikzset{%
  block/.style    = {draw, thick, rectangle, minimum height = 2.4em, minimum width = 2.6em},
  wideblock/.style    = {draw, thick, rectangle, minimum height = 2em, minimum width = 3.35em},
  box/.style    = {draw=none}, 
  sum/.style      = {draw, circle, minimum size=1.5em, inner sep = 0}, 
  input/.style    = {coordinate}, 
  output/.style   = {coordinate} 
}
\tikzset{
  -|-/.style={
    to path={
      (\tikztostart) -| ($(\tikztostart)!#1!(\tikztotarget)$) |- (\tikztotarget)
      \tikztonodes
    }
  },
  -|-/.default=0.5,
  |-|/.style={
    to path={
      (\tikztostart) |- ($(\tikztostart)!#1!(\tikztotarget)$) -| (\tikztotarget)
      \tikztonodes
    }
  },
  |-|/.default=0.5,
}
\tikzset{microphone/.style={black,circle,draw,fill=white,scale=1}}
\tikzset{conjunction/.style={black,circle,draw,fill=black,scale=0.25}}
\def\numChan#1#2#3{
\begin{scope}
	\node (rect) at #1 [draw=none, fill=none, minimum width=0.1*#2, minimum height=#2, inner sep = 0pt](#3){};
    \draw [<->,>=triangle 90 cap, thin](#3.south west) -- (#3.north east);
\end{scope}
}
\def\annotateAligned#1#2{
\begin{scope}
	\node (rect) at (#1.south west) [xshift = 0.1em, yshift = -0.3em, draw=none, fill=none, inner sep = 0pt, anchor = north west]{#2};
\end{scope}
}
\def\annotateCenter#1#2{
\begin{scope}
	\node (rect) at (#1.south) [yshift = -0.3em, draw=none, fill=none, inner sep = 0pt, anchor = north]{#2};
\end{scope}
}
\newlength\fheight 
\newlength\fwidth 
\pgfdeclarelayer{bg}    
\pgfsetlayers{bg,main}  

\graphicspath{{./fig/}}

\title{Integrated Sidelobe Cancellation and Linear Prediction Kalman Filter for Joint Multi- Microphone Speech Dereverberation, Interfering Speech Cancellation, and Noise Reduction}

\author{Thomas~Dietzen,~
        Simon~Doclo,~\IEEEmembership{Senior Member,}
        Marc~Moonen,~\IEEEmembership{Fellow,}
        and~Toon~van~Waterschoot~\IEEEmembership{Member} 
              
\thanks{T. Dietzen and T. van Waterschoot are with KU Leuven, Dept. of Electrical Engineering (ESAT), STADIUS Center for Dynamical Systems, Signal Processing and Data Analytics and ETC Technology Cluster Electrical Engineering, Leuven, Belgium. S. Doclo is with University of Oldenburg, Dept. of Medical Physics and Acoustics and the Cluster of Excellence Hearing4all, Oldenburg, Germany. M. Moonen is with KU Leuven, ESAT, STADIUS.}}      

\begin{document}

\maketitle

\begin{abstract}
In multi-microphone speech enhancement, reverberation as well as additive noise and/or interfering speech are commonly suppressed by deconvolution and spatial filtering, e.g., using multi-channel linear prediction (MCLP) on the one hand and beamforming, e.g., a generalized sidelobe canceler (GSC), on the other hand.
In this paper, we consider several reverberant speech components, whereof some are to be dereverberated and others to be canceled, as well as a diffuse (e.g., babble) noise component to be suppressed. 
In order to perform both deconvolution and spatial filtering, we integrate MCLP and the GSC into a novel architecture referred to as integrated sidelobe cancellation and linear prediction (ISCLP), where the sidelobe-cancellation (SC) filter and the linear prediction (LP) filter operate in parallel, but on different microphone signal frames.
Within ISCLP, we estimate both filters jointly by means of a single Kalman filter.
We further propose a spectral Wiener gain post-processor, which is shown to relate to the Kalman filter's posterior state estimate.
The presented ISCLP Kalman filter is benchmarked against two state-of-the-art approaches, namely first a pair of alternating Kalman filters respectively performing dereverberation and noise reduction, and second an MCLP+GSC Kalman filter cascade.
While the ISCLP Kalman filter is roughly $M^2$ times less expensive than both reference algorithms, where $M$ denotes the number of microphones, it is shown to perform 
similarly as compared to the former, and to outperform the latter.
\end{abstract}
\section{Introduction}
\label{sec:intro}

\IEEEPARstart{I}{n} many wide-spread speech processing applications such as hands-free telephony and distant automatic speech recognition, reverberation as well as additive noise and/or interfering speech impinging on a microphone may deteriorate the quality and intelligibility of the speech recordings \cite{beutelmann06}.
The demanding tasks of dereverberation, noise reduction and/or interfering speech cancellation, and in particular the conjunction of these therefore remain a subject of ongoing research, with multi-microphone-based approaches exploiting spatial diversity receiving particular interest \cite{two-stage-beam, schwartz15c, schwartz15b, delcroix07b, nakatani08, yoshioka09, yoshioka11, yoshioka12, yoshioka13, jukic15, delcroix15, yoshioka15, dietzen16b, braun16, jukic2017, dietzen17, BraunJune2018, heymann18, dietzen18, dietzen19TASLP, nakatani2019}.
In this context, we below briefly discuss two broad concepts in multi-microphone speech enhancement, namely spatial filtering and deconvolution.

As a spatial filtering technique, beamforming is commonly used in noise reduction and interfering speech cancellation, but may as well be applied for dereverberation \cite{two-stage-beam, schwartz15c, schwartz15b}.
In order to perform both dereverberation and noise reduction, several beamforming schemes have been proposed.
In \cite{two-stage-beam}, a cascaded approach is presented, using data-independent, super-directive beamforming for dereverberation, and data-dependent, e.g., minimum-variance distortionless response (MVDR) beamforming, for noise reduction.
The generalized sidelobe canceler (GSC), a popular implementation of the MVDR beamformer, has been applied in different constellations \cite{schwartz15c, schwartz15b}. 
In \cite{schwartz15c}, joint dereverberation and noise reduction is performed using a single GSC, while in \cite{schwartz15b}, a nested structure is proposed, employing an inner GSC for dereverberation and an outer GSC for noise reduction.
The GSC is composed of two parallel signal paths: a reference path and a sidelobe-cancellation (SC) path.
The reference path traditionally employs a matched filter (MF), while the SC path cascades a blocking matrix (BM), blocking either the entire or the early-reverberant speech component, and an SC filter, minimizing the output power and thereby suppressing residual nuisance components in the reference path, i.e. either residual noise or both residual noise and reverberation components.

As a deconvolution technique, multi-channel linear prediction (MCLP) \cite{delcroix07b, nakatani08, yoshioka09, yoshioka11, yoshioka12, yoshioka13, jukic15, delcroix15, yoshioka15, dietzen16b, braun16, jukic2017, dietzen17, BraunJune2018, heymann18, dietzen18, dietzen19TASLP, nakatani2019} recently prevailed in blind speech dereverberation, while noise reduction is not targeted.
As opposed to beamforming, MCLP does not require spatial information on the speech source. 
Instead, for each microphone, the reverberation component to be canceled is modeled as a linear prediction (LP) component, i.e. as a filtered version of the delayed microphone signals, with the LP filter to be estimated.
Besides iterative LP filter estimation approaches such as \cite{nakatani08, yoshioka11, yoshioka12, jukic15, delcroix15, yoshioka15}, also adaptive approaches based on recursive least squares (RLS) \cite{yoshioka09, yoshioka13, jukic2017, heymann18} as well as the Kalman filter \cite{dietzen16b, braun16, dietzen17} have been proposed in the past years.
In order to reduce noise after dereverberation, multiple-output MCLP has been cascaded with MVDR beamforming in \cite{delcroix15, yoshioka15}, which was seen to be a commonly adopted approach in the 2018 CHiME-5 challenge \cite{chime5}.
In \cite{nakatani2019}, the cascade in \cite{delcroix15, yoshioka15} is unified. 
In \cite{BraunJune2018}, joint MCLP-based dereverberation and noise reduction is performed using a pair of alternating Kalman filters respectively estimating the LP filter and the noise-free reverberant speech component.

In \cite{dietzen19TASLP}, we have presented a comparative analysis of the GSC and MCLP. 
In another previous paper \cite{dietzen18}, instead of cascading MCLP and beamforming or relying on beamforming only, we have proposed to integrate the GSC and MCLP by employing an SC path and LP path in parallel, resulting in an architecture we refer to as integrated sidelobe cancellation and linear prediction (\mbox{ISCLP}).
Within this novel architecture, we have estimated the SC and LP filters jointly by means of a single Kalman filter.  
Here, the spatial pre-processing blocks MF and BM require an estimate of the relative early transfer functions (RETFs), cf. also \cite{schwartz15c}, while the Kalman filter requires an estimate of the power spectral density (PSD) of the desired early speech component, cf. also \cite{dietzen16b, braun16, dietzen17}.
In this paper, the work in \cite{dietzen18} is extended in the following manner.
We generalize the short-time Fourier transform (STFT) domain-based signal model, which now comprises several (spatio-temporally correlated) reverberant speech components, whereof some are to be dereverberated and others to be canceled, as well as a (spatially, but not temporally correlated) diffuse (e.g., babble) noise component to be suppressed.
This generalized acoustic scenario necessitates (non-stationary) multi-source early PSD estimation and RETF updates, which is achieved by means of the algorithm recently proposed in \cite{dietzen19TBA}. 
We further augment the proposed approach by a spectral Wiener gain post-processor, which is shown to relate to the Kalman filter's posterior state estimate.
In order to demonstrate the effectiveness of the {ISCLP} Kalman filter, we compare against two state-of-the-art approaches -- first the previously mentioned alternating Kalman filters in \cite{BraunJune2018}, and second a MCLP+GSC Kalman filter cascade, conceptually relating to \cite{delcroix15, yoshioka15}. 
As compared to these two reference algorithms, the {ISCLP} Kalman filter is computationally roughly $M^2$ times less expensive, where $M$ denotes the number of microphones. 
Yet, the ISCLP Kalman filter is shown to perform similarly as compared to the alternating Kalman filters, and to outperform the MCLP+GSC Kalman filter cascade.
A MATLAB implementation is available at \cite{taslp19bCodeAudio}.

The paper is organized as follows.
In Sec. \ref{sec:signal}, we present the signal model in the STFT domain.
In Sec. \ref{sec:jointdereverb}, the ISCLP Kalman filter is described.
Implementational aspects are discussed in Sec. \ref{sec:implementationalaspects}, followed by simulations in Sec. \ref{sec:sim}.

\section{Signal Model}
\label{sec:signal}
Throughout the paper, we use the following notation: vectors are denoted by lower-case boldface letters, matrices by upper-case boldface letters, $\mathbf{I}$ and $\mathbf{0}$ denote an identity and zero matrix, $\mathbf{1}$ denotes a vector of ones, $\mathbf{A}^*$, $\mathbf{A}^\transp$, $\mathbf{A}^\herm$, and $\Exp[\mathbf{A}]$ denote the complex conjugate, the transpose, the complex conjugate transpose or Hermitian, and the expected value of a matrix $\mathbf{A}$.
The operation $\Diag[\a]$ creates a diagonal matrix with the elements of $\a$ on its diagonal, and $\tr[\A]$ denotes the trace of $\A$.
Submatrices  are referenced either by index ranges or alternatively by sets of indices, 
e.g., the submatrix of $\A$ spanning all rows and the columns $j_1$ to $j_2$ is denoted as $[\A]_{:,j_1:j_2}$, and the submatrix composed of all rows and the columns of $\A$ with indices in the ordered set $T$ is denoted as $[\A]_{:,\in T}$.

In the short-time Fourier transform (STFT) domain, with $l$ and $k$ indexing the frame and  the frequency bin, respectively, let $y_m(l,k)$ with $m=1,\dots,M$ denote the $m^\text{th}$ microphone signal, with $M$ the number of microphones.
In the following, we treat all frequency bins independently and hence omit the frequency index.
We define the stacked microphone signal vector $\y(l) \in \mathbb{C}^{M}$,
\begin{align}
\y(l) &= 
\begin{pmatrix}
y_1(l) & \cdots & y_{M}(l)
\end{pmatrix}^{\transp} 
\label{eq:sm:y_stacked}
\end{align} 
composed of the mutually uncorrelated reverberant speech components $\x_n(l)$ with $n=1,\dots,N$ originating from $N < M$ point speech sources and the noise component $\v(l)$, defined similarly to (\ref{eq:sm:y_stacked}), i.e.
\begin{align}
{\y}(l) &= \sum_{n=1}^N{\x}_n(l) + {\v}(l). 
\label{eq:sm:y_decomp}
\end{align}
Here, the reverberant speech components $\x_n(l)$ may be decomposed into the early and late-reverberant speech components ${\x}_{n|\textsl{e}}(l)$ and ${\x}_{n|\ell}(l)$, i.e.
\begin{align}
{\x}_n(l) &= {\x}_{n|\textsl{e}}(l)+ {\x}_{n|\ell}(l),
\label{eq:sm:xn_decomp}
\end{align}
which are commonly parted by the arrival time of the therein contained reflections
and assumed to have distinct spatio-temporal properties as outlined below.
Let ${\x}_{\textsl{e}}(l) =  \sum_{n=1}^N {\x}_{n|\textsl{e}}(l)$ and ${\x}_{\ell}(l) = \sum_{n=1}^N {\x}_{n|\ell}(l)$  denote the sum of the early and late-reverberant speech components, respectively, such that $\y(l)$ in (\ref{eq:sm:y_decomp})--(\ref{eq:sm:xn_decomp}) may alternatively be written as
\begin{align}
\y(l) = {\x}_{\textsl{e}}(l) + {\x}_{\ell}(l) + \v(l). \label{eq:sm:y_rewritten}
\end{align}
Early reflections are assumed to arrive within the same frame, where the early components in ${\x}_{n|\textsl{e}}(l)$ are related by the RETFs in $\h_n(l) \in \mathbb{C}^{M}$ as ${\x}_{n|\textsl{e}}(l) =  \h_n(l)s_{n}(l)$. Here, without loss of generality, the RETFs are assumed to be relative to the first microphone, i.e. $[\h_n(l)]_1 = 1$, and $s_{n}(l) = [{\x}_{n|\textsl{e}}(l)]_1$ denotes the early component in the first microphone originating from the $n^{\text{th}}$ source, in the following referred to as early source image.
We stack $\h_n(l)$ and $s_{n}(l)$ into ${\H}(l) \in \mathbb{C}^{M \times N}$ and $\mathbf{s}(l) \in  \mathbb{C}^{N}$, respectively, i.e.
\begin{align}
 {\H}(l) & = 
 \begin{pmatrix}
 \h_1(l) & \cdots &\h_N(l)
 \end{pmatrix},
 \label{eq:sm:H_stacked}\\ 
\mathbf{s}(l) &= 
  \begin{pmatrix}
  s_{1}(l) &\cdots &s_{N}(l)
  \end{pmatrix}^\transp, 
  \label{eq:sm:xne_stacked}
\end{align}
such that ${\x}_{\textsl{e}}(l)$ is expressed by
\begin{align}
{\x}_{\textsl{e}}(l) =  {\H}(l)\mathbf{s}(l). 
 \label{eq:sm:sumxne_matmult}
\end{align}
In the following, let $N_T \leq N$ early speech source images $s_n(l)$ be defined as the target source images,
and let $T$ denote the set of the corresponding $\vert T\vert = N_T$ target-source indices.
Let $T'$ denote the complement set to $T$, with $\vert T'\vert = N-N_T$.
In order to distinguish the target components in $\y(l)$ as well as their complements, we introduce the short-hand notations similar to (\ref{eq:sm:H_stacked})--(\ref{eq:sm:sumxne_matmult}),
\begin{align}
\H_{T}(l) &= [\H(l)]_{:,\in T}, \label{eq:sm:HT}\\
\mathbf{s}_{T}(l) &= [\mathbf{s}(l)]_{\in T}, \label{eq:sm:sT}\\[3pt]
\x_{\textsl{e}|T}(l) &= \H_{T}(l)\mathbf{s}_{T}(l),\label{eq:sm:xeT}
\end{align}
and $\H_{T'}(l)$, $\mathbf{s}_{T'}(l)$, and $\x_{\textsl{e}|T'}(l)$ similarly, such that ${\x}_{\textsl{e}}(l)$ in (\ref{eq:sm:y_rewritten}) becomes 
\begin{align}
{\x}_{\textsl{e}}(l) = \x_{\textsl{e}|T}(l) + \x_{\textsl{e}|T'}(l).\label{eq:sm:xe_split}
\end{align}
Our objective is to estimate
\begin{align}
s_{T}(l) = \sum_{n\in T} s_n(l) = \mathbf{1}^\transp\mathbf{s}_{T}(l)\label{eq:sm:st}
\end{align}
from $\y(l)$ by means of the ISCLP Kalman filter.
To this end, we rely on assumptions on the spatio-temporal behavior of the individual microphone signal components. 
We assume that $s_n(l)$ is temporally uncorrelated across frames, i.e. we have $\Exp [s_n(l\-l') s_n^*(l)] = 0$  for $l'>0$, and with  ${\x}_{n|\textsl{e}}(l) =  \h_n(l)s_{n}(l)$ consequently  
\begin{align}
\Exp [\x_{n|\textsl{e}}(l\-l') \x_{n|\textsl{e}}^\herm(l)] = \mathbf{0} \quad \text{for}\quad l'>0.  \label{eq:xe_uncorrelated}
\end{align}
For speech signals, this assumption can be considered approximately justified if the STFT window length and window shift are sufficiently large.
Within the limits defined by the reverberation time, we assume that the late-reverberant speech component $\x_{n|\ell}(l)$ is correlated to previous early source images  $s_n(l\-l')$ with $l'>0$, but not to the current early source image $s_n(l)$, i.e. we have
\begin{align}
\Aboxed{
\Exp [\x_{n|\textsl{e}}(l\-l') \x_{n|\ell}^\herm(l)] 
&\neq \mathbf{0}  \quad \text{for}\quad l'>0,}\label{eq:xell_correlated}\\
\Exp [\x_{n|\textsl{e}}(l) \x_{n|\ell}^\herm(l)]  &= \mathbf{0}.\label{eq:xell_xl_uncorrelated}
\end{align}
Note that (\ref{eq:xell_correlated}) also implies $\Exp [\x_{n|\ell}(l\-l') \x_{n|\ell}^\herm(l)] \neq \mathbf{0}$ for all $l'$.
With  (\ref{eq:xell_correlated}), we may predict $\x_{n|\ell}(l)$ from $\x_n(l\-l')$, which indeed is the fundamental assumption of MCLP-based dereverberation \cite{delcroix07b, nakatani08, yoshioka09, yoshioka11, yoshioka12, yoshioka13, jukic15, delcroix15, yoshioka15, dietzen16b, braun16, jukic2017, dietzen17, BraunJune2018, heymann18}.
Assumptions (\ref{eq:xe_uncorrelated}) and (\ref{eq:xell_xl_uncorrelated}) allow for unbiased filter estimation \cite{dietzen19TASLP} in MCLP-based dereverberation \cite{delcroix07b, nakatani08, yoshioka09, yoshioka11, yoshioka12, yoshioka13, jukic15, delcroix15, yoshioka15, dietzen16b, braun16, jukic2017, dietzen17, BraunJune2018, heymann18} and GSC-based dereverberation and noise reduction \cite{schwartz15c, schwartz15b}, respectively.
Hence, all three assumptions  (\ref{eq:xe_uncorrelated})--(\ref{eq:xell_xl_uncorrelated})  are equally essential in the derivation of the ISCLP Kalman filter, cf. Sec. \ref{sec:jointdereverb}.
Similarly to $s_n(l)$, the noise component $\v(l)$ is assumed to be temporally uncorrelated, i.e. 
\begin{align}
\Exp [\v(l\-l') \v^\herm(l)] = \mathbf{0} \quad \text{for}\quad l'>0,  \label{eq:b_uncorrelated}
\end{align}
and is therefore not predictable.

Within frame $l$, i.e. for $l'=0$, we further make assumptions on the spatial behavior of $\x_{n|\ell}(l)$ and $\v(l)$, namely that both may be modeled as spatially diffuse.
However, as these assumptions are irrelevant in the derivation of the ISCLP Kalman filter itself, cf. Sec. \ref{sec:jointdereverb}, 
but required only for parameter estimation based on \cite{dietzen19TBA}, i.e. the estimation of the RETFs $\H_T(l)$ and the PSD  $\varphi_{s_{T}}(l) = \Exp[s_{T}(l)s_{T}^*(l)]$, we treat them in the corresponding section only, cf. Sec. \ref{sec:targetPSDRETFupdate}.

\section{Integrated Sidelobe Cancellation\\and Linear Prediction Kalman Filter}
\label{sec:jointdereverb}

We strive to estimate the target component $s_T(l)$ from the microphone signals $\y(l)$ defined in Sec. \ref{sec:signal}.
For this purpose, we introduce the \mbox{ISCLP} architecture.
In Sec. \ref{sec:ISCLP}, we describe the SC and LP signal paths and filter constellations, 
which require spatio-temporal pre-processing of $\y(l)$.
In Sec. \ref{sec:filterestim}, striving for recursive filter estimation,
we define an ISCLP state-space model for the SC and the LP filter, whereof a Kalman filter is deduced. 
The Kalman filter yields a (prior) estimate $e(l) = \hat{s}_T(l)$ of $s_T(l)$, which may further be spectrally post-processed, as shown 
in Sec. \ref{sec:spectralpost}.

\subsection{ISCLP Signal Path Architecture}
\label{sec:ISCLP}

A block-diagram of the \mbox{ISCLP} architecture is depicted in Fig. \ref{fig:ISCLP}.
It integrates the GSC and MCLP and hence consists of three signal paths: a reference path employing an MF, an SC path, composed of a BM and an SC filter, and a LP path, composed of a delay and an LP filter.
While the MF, the BM and the SC filter are multiplicative (mult.), i.e. they operate on a single frame, the LP filter is convolutive (conv.), i.e. it operates across frames.
The MF and the BM perform spatial pre-processing, serving unconstrained estimation of the SC filter, while the delay may analogously be considered as temporal pre-processing, serving unconstrained estimation of the LP filter.
Structurally, one may interpret \mbox{ISCLP} either as MCLP with the conventional reference channel selection replaced by a GSC, or alternatively as a GSC employing a generalized BM (composed of a traditional BM and a delay line), and a convolutive filter (composed of the SC and the LP filter).
In the following, we formally discuss the individual signal paths.

In oder to maintain the target component $s_{T}(l)$ in (\ref{eq:sm:st}), the MF $\g \in \mathbb{C}^{M}$ must satisfy \cite{doclo2010acoustic, gannot17} 
\begin{align}
\g^\herm(l)\H_T(l) = \mathbf{1}^\transp, \label{eq:isclp:path:gconstraint}
\end{align}
where a commonly used \cite{doclo2010acoustic, gannot17} choice for $\g(l)$ adhering to (\ref{eq:isclp:path:gconstraint}) is  
\begin{align}
\g(l) &=  \H_T(l)\bigl(\H_T^\herm(l)\H_T(l)\bigr)^\inv \mathbf{1}, \label{eq:g}
\end{align}
with $\H_T(l)\bigl(\H_T^\herm(l)\H_T(l)\bigr)^\inv$ the pseudoinverse of $\H^\herm_T(l)$.
In practice, we hence require an estimate $\hat{\H}_T(l)$ of $\H_T(l)$, cf. also Sec. \ref{sec:targetPSDRETFupdate}.
With $\y(l)$ as in (\ref{eq:sm:y_rewritten}), combining (\ref{eq:sm:xeT})--(\ref{eq:sm:st}), 
the MF output $q(l)$ becomes
\begin{align}
\Aboxed{
q(l) &= \g^\herm(l)\y(l)}\nonumber\\
&= s_{T}(l) + \g^\herm(l)\bigl(\x_{\textsl{e}|T'}(l) + \x_{\ell}(l) + \v(l)\bigr). \label{eq:q}
\end{align}
The BM $\B(l) \in \mathbb{C}^{M \times M-N_T}$ must be orthogonal to $\H_T(l)$, i.e. 
\begin{align}
{\mathbf{B}}^\herm(l) \H_T(l) = \mathbf{0},
\end{align}
and with (\ref{eq:g}) hence $\B^\herm(l)\g(l) = \mathbf{0}$. 
One may, e.g., choose ${\mathbf{B}}(l)$ based on the first $M-N_T$ columns of the rank-$(M\-N_T)$ projection matrix to the null space of $\H_T(l)$ \cite{gannot17}, i.e.
\begin{align}
{\mathbf{B}}(l) &= \bigl[\I - \H_T(l)\bigl(\H_T^\herm(l)\H_T(l)\bigr)^\inv \H_T^\herm(l)\bigr]_{:,1:M-N_T},\label{eq:B}
\end{align}
with $\H_T(l)\bigl(\H_T^\herm(l)\H_T(l)\bigr)^\inv \H_T^\herm(l)$ the projection matrix to the column space of $\H_T(l)$.
With $\y(l)$ as in (\ref{eq:sm:y_rewritten}), combining (\ref{eq:sm:xeT})--(\ref{eq:sm:xe_split}), the SC-filter input $\u_\textsl{\tiny{SC}}(l) \in \mathbb{C}^{M-N_T}$ is then given by
\begin{align}
\Aboxed{
\u_\textsl{\tiny{SC}}(l) &= \B^\herm(l)\y(l)}\nonumber\\
&= \B^\herm(l)\bigl(\x_{\textsl{e}|T'}(l) + \x_\ell(l) + \v(l)\bigr), \label{eq:uSC}
\end{align}
whereby the target component $\x_{\textsl{e}|T}(l) = \H_T(l)\mathbf{s}_T(l)$ is canceled.
Using a delay of one\footnote{
In MCLP literature, delays of more than one frame are commonly used \cite{yoshioka11, yoshioka12, yoshioka13, jukic15, delcroix15, yoshioka15, braun16, jukic2017, BraunJune2018, heymann18} in order to  avoid temporal target component leakage due to overlapping windows in the STFT processing, cf. Sec. \ref{sec:leakage}. 
As we here also consider interfering reverberant speech components to be canceled, larger delays in the LP filter path however call for a convolutive SC filter
 \cite{dietzen19TASLP} instead. 
The here proposed design did not show to be sensitive to leakage effects, cf. Sec. \ref{sec:leakage} and Sec. \ref{sec:sim}.
}  frame, the LP-filter input  $\u_\textsl{\tiny{LP}}(l) \in \mathbb{C}^{(L-1)M}$ is defined by stacking $\y(l)$ over the past $L-1$ frames, i.e.
\begin{align}
\Aboxed{
\u_\textsl{\tiny{LP}}(l) &= 
{\begin{pmatrix}
\y^\transp(l\-1) & \cdots & \y^\transp(l\-L\+1)
\end{pmatrix}}^\transp. }\label{eq:uLP}
\end{align}
With the SC filter $\hat{\w}_\textsl{\tiny{SC}}(l)  \in \mathbb{C}^{M-N_T}$ and the LP filter $\hat{\w}_\textsl{\tiny{LP}}(l)  \in \mathbb{C}^{(L-1)M}$, the enhanced signal $e(l) = \hat{s}_T(l)$ at the output of \mbox{ISCLP}, also referred to as error signal in the remainder, is given by
\begin{align}
e(l) &= \hat{s}_T(l) = q(l) - z_\textsl{\tiny{SC}}(l) - z_\textsl{\tiny{LP}}(l), \label{eq:e}\\
\text{with} \quad z_\textsl{\tiny{SC}}(l) &= \hat{\w}_\textsl{\tiny{SC}}^\herm(l)\u_\textsl{\tiny{SC}}(l), \label{eq:zSC}\\
z_\textsl{\tiny{LP}}(l) &= \hat{\w}_\textsl{\tiny{LP}}^\herm(l)\u_\textsl{\tiny{LP}}(l). \label{eq:zLP}
\end{align}
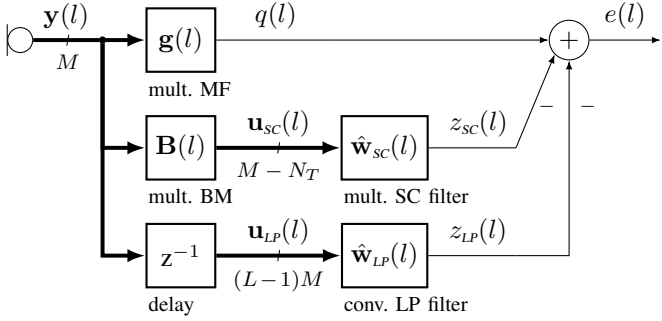
\begin{figure}[!t]
\centering
\begin{tikzpicture}[auto, node distance=6em, >=triangle 45]
\draw
	node at (0,0)[microphone, name=source] {} 
	node [conjunction, right =2.5em of source] (conj) {}
	node [block, right = 1.5em of conj] (beamformer) {$\mathbf{g}(l)$}
	node [block, below =1.6em of beamformer] (blockingmatrix) {$\mathbf{B}(l)$}
	node [block, below =1.6em of blockingmatrix] (delay) {$\text{z}^{-1}$}
    node [sum, right=19.5em of source] (sum) {$+$}
    node at ($(sum)+(-.85em,-2.5em)$){$\scriptstyle -$}
	node at ($(sum)+(+.7em,-2.5em)$){$\scriptstyle -$}
	node [block, right=4.7em of blockingmatrix] (SCfilter) {$\hat{\w}_\textsl{\tiny{SC}}(l)$}
	node [output, right =3.3em of SCfilter] (conj2) {}
	node [block, below =1.6em of SCfilter] (LPfilter) {$\hat{\w}_\textsl{\tiny{LP}}(l)$}
	node [output, right =3.3em of LPfilter] (conj3) {}
	node [output, right=2.7em of sum] (target){}
    node [output, below =4em of sum] (conj4) {}
    node [output, right =4.5em of beamformer] (conj5) {};

    \draw ([yshift=6pt]source.west) -- ([yshift=-6pt]source.west);
    \draw[-,line width=1.5pt](source) -- node[yshift=.0em]{$\mathbf{y}(l)$}(conj);
    \draw[-{Latex[scale=.7]},line width=1.5pt](conj) |- node{}(blockingmatrix);
    \draw[-{Latex[scale=.7]},line width=1.5pt](conj) |- node{}(delay);
    \draw[-{Latex[scale=.7]},line width=1.5pt](blockingmatrix) -- node[yshift=.0em]{$\mathbf{u}_\textsl{\tiny{SC}}(l)$}(SCfilter);
    \draw[-{Latex[scale=.7]},line width=1.5pt](delay) -- node[yshift=.0em]{$\mathbf{u}_\textsl{\tiny{LP}}(l)$}(LPfilter);
    \draw[-{Latex[scale=.7]},line width=1.5pt](conj) -- node{}(beamformer);
    \draw[-](beamformer) -- node[yshift=.07em]{$q(l)$}(conj5);
	\draw[-{Latex[scale=.7]}](conj5) -- (sum);
    
    
    \draw[-](SCfilter) -- node[xshift=.23em, yshift=.07em]{$z_\textsl{\tiny{SC}}(l)$}(conj2);
    \draw[-{Latex[scale=.7]}](conj2) -- (sum.south west);
    
    \draw[-](LPfilter) -- node[xshift=.23em, yshift=.07em]{$z_\textsl{\tiny{LP}}(l)$}(conj3);
    \draw[-{Latex[scale=.7]}](conj3) -| (sum.south);
    
    \draw[-{Latex[scale=.7]}](sum) -- node[yshift=.07em]{$e(l)$}(target);
	\numChan{($(source)+(1.8em,0)$)}{4pt}{numChan1};
    \numChan{($(blockingmatrix)+(3.7em,0)$)}{4pt}{numChan2};
    \numChan{($(delay)+(3.7em,0)$)}{4pt}{numChan3};

    \annotateAligned{beamformer}{\footnotesize{mult. MF}};
    \annotateAligned{blockingmatrix}{\footnotesize{mult. BM}};
    \annotateAligned{delay}{\footnotesize{delay}};
    \annotateAligned{SCfilter}{\footnotesize{mult. SC filter}};
    \annotateAligned{LPfilter}{\footnotesize{conv. LP filter}};
	\annotateCenter{numChan1}{\footnotesize{$M$}};
	\annotateCenter{numChan2}{\footnotesize{$M-N_T$}};
	\annotateCenter{numChan3}{\footnotesize{$(L\hskip 0.13em{-}\hskip 0.13em1)M$}};

\end{tikzpicture}
\caption{The integrated sidelobe cancellation and linear prediction (\mbox{ISCLP}) architecture.}
\label{fig:ISCLP}
\end{figure}
At this point, given $q(l)$, $\u_\textsl{\tiny{SC}}(l)$, and $\u_\textsl{\tiny{LP}}(l)$, our task consists in obtaining the filters $\hat{\w}_\textsl{\tiny{SC}}(l)$ and $\hat{\w}_\textsl{\tiny{LP}}(l)$ as estimates of some yet to be defined associated true states $\w_\textsl{\tiny{SC}}(l)$ and $\w_\textsl{\tiny{LP}}(l)$, cf. Sec \ref{sec:filterestim}.
In this respect, let us first discuss the mutual relations between the target component  $s_{T}(l)$ in $q(l)$ and the signals $\u_\textsl{\tiny{SC}}(l)$ and $\u_\textsl{\tiny{LP}}(l)$, as well as the consequences thereof for the filter estimation.
Note that due to the delay in the LP path, the filter estimates $\hat{\w}_\textsl{\tiny{SC}}(l)$ and $\hat{\w}_\textsl{\tiny{LP}}(l)$ do not operate on the same input-data frame at the same time.
The SC-filter input $\u_\textsl{\tiny{SC}}(l)$ in (\ref{eq:uSC}) depends on the current frame $\y(l)$ only, such that $\hat{\w}_\textsl{\tiny{SC}}(l)$ will exploit spatial correlations within the current frame.
Due to the cancellation of $\x_{\textsl{e}|T}(l)$ at the BM output and (\ref{eq:xell_xl_uncorrelated}), we have $\Exp [\u_\textsl{\tiny{SC}}(l) s_{T}^*(l)] = \mathbf{0}$. 
This allows for unconstrained, recursive estimation of ${\w}_\textsl{\tiny{SC}}(l)$, which is indeed the general incentive behind the usage of  GSC-like structures \cite{doclo2010acoustic, gannot17}. 
In contrast, the LP-filter input $\u_\textsl{\tiny{LP}}(l)$ in (\ref{eq:uLP}) depends on the $L\-1$ previous frames  $\y(l\-l')$ with $l'=1,\dots,L\-1$, such that $\hat{\w}_\textsl{\tiny{LP}}(l)$ will exploit spatio-temporal correlations between the current and the previous frames (but not within the current frame).
Due to this delay and (\ref{eq:xe_uncorrelated}), we have $\Exp [\u_\textsl{\tiny{LP}}(l) s_{T}^*(l)] = \mathbf{0}$, likewise allowing for unconstrained, recursive estimation of ${\w}_\textsl{\tiny{LP}}(l)$.
However, with both $\u_\textsl{\tiny{SC}}(l)$ and $\u_\textsl{\tiny{LP}}(l)$ containing (late-)reverberant components, the two inputs are \textit{not} independent, i.e. 
\begin{align}
\Aboxed{
\Exp [\u_\textsl{\tiny{LP}}(l)\u_\textsl{\tiny{SC}}^\herm(l)] \neq \mathbf{0}, }
\end{align}
cf. (\ref{eq:xell_correlated}), and as a consequence also $\Exp [z_\textsl{\tiny{LP}}(l) z_\textsl{\tiny{SC}}^*(l)] \neq 0$.
In other words, a change in $\hat{\w}_\textsl{\tiny{SC}}(l)$ requires a change in $\hat{\w}_\textsl{\tiny{LP}}(l)$, and vice versa. 
We therefore strive to \textit{jointly} estimate both filters.

\subsection{ISCLP State-Space Model and Kalman Filter Update}
\label{sec:filterestim}

In order to recursively estimate the SC and LP filter, we employ a Kalman filter, which has also been applied successfully to MCLP in previous works \cite{dietzen16b, braun16, dietzen17, BraunJune2018}.
Hereby, we interpret $\hat{\w}_\textsl{\tiny{SC}}(l)$ and $\hat{\w}_\textsl{\tiny{LP}}(l)$ as estimates of the true states $\w_\textsl{\tiny{SC}}(l)$ and $\w_\textsl{\tiny{LP}}(l)$, which are defined by a state-space model comprising the so-called measurement equation and the process equation.
 In the following, we first define the state-space model, and then present the corresponding Kalman filter update equations, which recursively estimate the true state.

As we intend to estimate $\w_\textsl{\tiny{SC}}(l)$ and $\w_\textsl{\tiny{LP}}(l)$ jointly, cf. Sec. \ref{sec:ISCLP}, we stack the SC and LP filter path into $\u(l) \in \mathbb{C}^{LM-N_T}$ and  $\w(l) \in \mathbb{C}^{LM-N_T}$, i.e.
\begin{align}
\u(l) = 
\begin{pmatrix}
\u_\textsl{\tiny{SC}}^\transp(l)  & \u_\textsl{\tiny{LP}}^\transp(l)
\end{pmatrix}^\transp ,\label{eq:ustacked}\\
\w(l) = 
\begin{pmatrix}
\w_\textsl{\tiny{SC}}^\transp(l)  & \w_\textsl{\tiny{LP}}^\transp(l)
\end{pmatrix}^\transp. \label{eq:wstacked}
\end{align}
and $\hat{\w}(l)$ defined similarly to (\ref{eq:wstacked}).
The true state ${\w}(l)$ is considered a random variable with zero mean and correlation matrix $\bPsi_{w}(l)  = \Exp[\w(l)\w^\herm(l)]$.
We assume that ${\w}(l)$ leads to complete cancellation\footnote{Note that complete cancellation may not necessarily be possible, e.g., if $\v(l) \neq \mathbf{0}$  \cite{dietzen19TASLP}, and so the true state does not necessarily exist.
Nonetheless, lacking deeper knowledge on the true system, we assume that it lies in the model set.
} 
 of $\g^\herm(l)\bigl(\x_{\textsl{e}|T'}(l) + \x_{\ell}(l) + \v(l)\bigr)$, and therefore yielding $e(l) = s_{T}(l)$, cf. (\ref{eq:q}) and (\ref{eq:e})--(\ref{eq:zLP}).
Reformulating (\ref{eq:e})--(\ref{eq:zLP}) using (\ref{eq:ustacked})--(\ref{eq:wstacked}), inserting $e(l) = s_{T}(l)$ and rearranging yields the so-called measurement equation,
\begin{align}
\Aboxed{q^*(l) &=  \u^\herm(l)\w(l) + s_{T}^*(l).} \label{eq:obsEq}
\end{align}
In Kalman filter terminology, we refer to $q^*(l)$ as the measurement and to  $s_{T}^*(l)$ as the (presumed zero-mean and temporally uncorrelated, cf. also Sec. \ref{sec:signal}) measurement noise with PSD $\varphi_{s_{T}}(l) = \Exp[s_{T}(l)s_{T}^*(l)]$. 
In practice, in order to implement the Kalman filter update equations, an estimate $\hat{\varphi}_{s_{T}}(l)$ of $\varphi_{s_{T}}(l)$ is required, cf. Sec. \ref{sec:targetPSDRETFupdate}.

The true state $\w(l)$ is assumed time-varying, which accounts for potential time variations in the room impulse responses, e.g., caused by time-varying source and microphone-array positions, as well as time-varying activity of individual sources and noise powers.
The so-called process equation models the evolution of the true state $\w(l)$ in the form of a first-order difference equation, i.e.
\begin{align}
\Aboxed{\w(l) &= \A^\herm(l)\w(l\-1) + \w_{\hspace{-.05em}\scriptscriptstyle{\Delta}}(l).}  \label{eq:procEq}
\end{align}
where $\A(l)$ models the state transition from one frame to the next, and the process noise $\w_{\hspace{-.04em}\scriptscriptstyle{\Delta}}(l)$ models a random (presumed zero-mean and temporally uncorrelated) variation component with correlation matrix $\bPsi_{w_{\hspace{-.05em}\scriptscriptstyle{\Delta}}}(l) = \Exp[\w_{\hspace{-.04em}\scriptscriptstyle{\Delta}}(l)\w^\herm_{\hspace{-.04em}\scriptscriptstyle{\Delta}}(l)]$.
Lacking deeper knowledge on the exact evolution of the true state, both $\A(l)$ and $\bPsi_{w_{\hspace{-.05em}\scriptscriptstyle{\Delta}}}(l)$ are commonly considered design parameters to be tuned \cite{enzner06, braun16, dietzen17, BraunJune2018}, 
cf. Sec. \ref{sec:initproc}.

The true state $\w(l)$ modeled by (\ref{eq:obsEq})--(\ref{eq:procEq}) may be estimated recursively by means of the Kalman filter update equations \cite{haykin02, simon2006optimal}, which are commonly presented as two distinct sets of updates per recursion, namely an a-priori time update reflecting the state evolution, cf. (\ref{eq:procEq}), and an a-posteriori measurement update reflecting the current measurement, cf. (\ref{eq:obsEq}).
Specifically, let $\hat{\w}(l)$ and $\hat{\w}^{+}(l)$ denote the yet to be defined prior and posterior state estimates of $\w(l)$, respectively, and let  $\tilde{\w}(l)$ and  $\tilde{\w}^{+}(l)$ denote the associated state estimation errors, i.e. 
\begin{align}
\tilde{\w}(l) &= \hat{\w}(l) - \w(l), \label{eq:wtilde}\\
\tilde{\w}^{+}(l) &= \hat{\w}^{+}(l) - \w(l), \label{eq:wtilde_post}
\end{align}
with the associated state estimation error correlation matrices $\bPsi_{\tilde{w}}(l)$ and $\bPsi^{+}_{\tilde{w}}(l)$.
Then, based upon (\ref{eq:procEq}) and (\ref{eq:obsEq}), respectively, the prior and posterior state estimates $\hat{\w}(l)$ and $\hat{\w}^{+}(l)$ shall recursively minimize the expected squared Euclidian norm of the associated state estimation error, i.e. 
 $\Exp\bigl[\Vert\tilde{\w}(l)\Vert^2\bigr] = \tr[\bPsi_{\tilde{w}}(l)]$ and $\Exp\bigl[\Vert\tilde{\w}^{+}(l)\Vert^2\bigr] = \tr[\bPsi_{\tilde{w}}^{+}(l)]$.
This leads to the celebrated Kalman filter update equations \cite{haykin02, simon2006optimal},
\begin{align}
\allowdisplaybreaks
\hat{\w}(l) 				&= \A^\herm(l)\hat{\w}^{+}(l\-1), 														\label{tuw}\\
\bPsi_{\tilde{w}}(l) 				&=  \A^\herm(l)\bPsi_{\tilde{w}}^+(l\-1)\A(l)+\bPsi_{w_{\hspace{-.04em}\scriptscriptstyle{\Delta}}}(l),		\label{tuP}\\[.25em] 
e^*(l) 			&=  q^*(l) -\u^\herm(l)\hat{\w}(l),															\label{E}\\
\varphi_{e}(l) 					&= \u^\herm(l)\bPsi_{\tilde{w}}(l)\u(l) + {\varphi}_{s_T}(l), 				\label{Psie}\\
\k(l) 							&= \bPsi_{\tilde{w}}(l)\u(l)\varphi_{e}^{-1}(l),											\label{K} \\[.25em] 
\hat{\w}^+(l) 			&= \hat{\w}(l) + \k(l)e^*(l), 																\label{muw}\\
\bPsi_{\tilde{w}}^+(l) 			&= \bPsi_{\tilde{w}}(l)-\k(l)\u^\herm(l)\bPsi_{\tilde{w}}(l), 							 			\label{muP}
\end{align}
where the time and the measurement update are given by (\ref{tuw})--(\ref{tuP}) and (\ref{muw})--(\ref{muP}), respectively.
In the time update, cf. (\ref{tuw})--(\ref{tuP}), the previously acquired posterior quantities $\hat{\w}^{+}(l\-1)$ and $\bPsi_{\tilde{w}}^{+}(l\-1)$ are propagated according to the evolution of the state ${\w}(l)$, cf. (\ref{eq:procEq}), yielding the prior quantities  $\hat{\w}(l)$  and  $\bPsi_{\tilde{w}}(l)$.
Then, given $\hat{\w}(l)$ and $\bPsi_{\tilde{w}}(l)$, the complex conjugate error signal $e^*(l)$, its PSD $\varphi_{e}(l)$, and the Kalman gain $\k(l)$ are computed, cf. (\ref{E})--(\ref{K}), thereby leveraging new information in terms of the measurement $q^*(l)$ and the measurement noise PSD ${\varphi}_{s_T}(l)$, cf. (\ref{eq:obsEq}).
Finally, in the measurement update, cf. (\ref{muw})--(\ref{muP}), $e^*(l)$ and $\k(l)$  are utilized to update $\hat{\w}(l)$ and $\bPsi_{\tilde{w}}(l)$, yielding the posterior quantities  $\hat{\w}^{+}(l)$  and  $\bPsi^{+}_{\tilde{w}}(l)$.
The error signal $e(l)$ in (\ref{E}) thereby represents the Kalman filter estimate of ${s}_T(l)$, cf. also (\ref{eq:e})--(\ref{eq:zLP}).
As the Kalman filter minimizes $\tr[\bPsi_{\tilde{w}}(l)]$ during convergence, it is easily seen that also $\varphi_{e}(l) = \Exp\bigl[\vert e(l)\vert^2\bigr]$ in (\ref{E}) is minimized. 
The Kalman filter requires initialization, which we consider in Sec. \ref{sec:initproc}.

\subsection{Posterior-like Spectral Post-Processing}
\label{sec:spectralpost}

With $\hat{\w}(l)$ a prior estimate of $\w(l)$, we may consider $e(l) = \hat{s}_T(l)$ in ({\ref{E}}) a prior estimate of ${s}_T(l)$.
After the measurement update in (\ref{muw}), yielding the posterior estimate $\hat{\w}^+(l)$ of $\w(l)$, we may accordingly define a posterior estimate $e^{+}(l) =  \hat{s}_T^{+}(l)$ similar to ({\ref{E}}) by
\begin{align}
e^{*|+}(l) 								&=  \hat{s}_T^{*|+}(l) = q^*(l) -\u^\herm(l)\hat{\w}^{+}(l).															\label{E_post}
\end{align}
Interestingly, $e^{+}(l)$ in (\ref{E_post}) can be shown to be a spectrally post-processed version of $e(l)$.
Precisely, inserting (\ref{muw}) while using (\ref{E}), inserting (\ref{K}) and finally (\ref{Psie}), we find
\begin{align}
e^{+}(l) 	= \hat{s}_T^{+}(l)			&= \bigl(1-\u^\herm(l)\k(l)\bigr)^{*}e(l) 		\nonumber\\
												&= \bigl(1-\u^\herm(l)\bPsi_{\tilde{w}}(l)\u(l)\varphi_{e}^{-1}(l) \bigr)^{*}e(l) 	 \nonumber\\
												&= \frac{ {\varphi}_{s_T}(l) }{\varphi_{e}(l) }e(l),	\label{s_hat_post}
\end{align}
where $\gamma(l) = {\varphi}_{s_T}(l)/\varphi_{e}(l)$ can be recognized as the spectral Wiener gain minimizing $\Exp\bigl[\vert {s}_T(l)-\gamma(l)e(l)\vert^2\bigr]$.
In practice, where we rely on potentially highly non-stationary estimates $\hat{\varphi}_{s_{T}}(l)$, cf. Sec. \ref{sec:leakage} and Sec. \ref{sec:targetPSDRETFupdate}, 
one may prefer slowly decaying gains for perceptual reasons \cite{loizou2007speech}.
Therefore, instead of using (\ref{s_hat_post}), we propose to alternatively define $\gamma(l)$ and $e^{+}(l)$ by
\begin{align}
\gamma(l) &= \max\biggl[\frac{{\varphi}_{s_T}(l)}{\varphi_{e}(l)},\beta\gamma(l-1)\biggr], \label{alt_gain}\\
e^{+}(l) =  \hat{s}_T^{+}(l) &= \gamma(l)e(l), \label{e_post_altGain}
\end{align}
with the tuning parameter $\beta \in [0, 1]$ limiting the gain decay.
%
Note that 
  (\ref{alt_gain})--(\ref{e_post_altGain}) reduce to (\ref{s_hat_post}) for $\beta = 0$, and to (\ref{E}) for $\beta = 1$ and $\gamma(0) = 1$ as initial gain, since ${\varphi}_{s_T}(l)/{\varphi}_{e}(l) \leq 1$ due to (\ref{Psie}).

\section{Implementational Aspects}
\label{sec:implementationalaspects}

Kalman filters perform optimally if the assumed state-space model matches the true system \cite{haykin02, simon2006optimal}.
In a practical implementation, the here presented ISCLP Kalman filter derived from the ISCLP state-space model in (\ref{eq:obsEq})--(\ref{eq:procEq})
 is subject to modeling errors, requires parameter estimation, and, where deeper knowledge on the underlying system dynamics is not available, parameter tuning.
These implementational aspects are discussed in the following.
In Sec. \ref{sec:leakage}, we qualitatively discuss the potential target component leakage due to imperfect spatio-temporal pre-processing in ISCLP and its impact on the proposed ISCLP Kalman filter. 
In Sec. \ref{sec:targetPSDRETFupdate}, we summarize a recently proposed approach to early PSD estimation and recursive RETF updating, which we employ in conjunction with the Kalman filter.
In Sec. \ref{sec:initproc}, we discuss the process equation parameter tuning and Kalman filter initialization.

\subsection{Spatio-Temporal Target Component Leakage}
\label{sec:leakage}

The previously made assumptions that $\Exp [\u_\textsl{\tiny{SC}}(l) s_{T}^*(l)] = \mathbf{0}$ and $\Exp [\u_\textsl{\tiny{LP}}(l) s_{T}^*(l)] = \mathbf{0}$, cf. Sec. \ref{sec:ISCLP}, may not be strictly satisfied in a practical implementation, which we refer to as \textit{target component leakage}.
Leakage may occur due to the following reasons. 
The spatial pre-processing components MF and BM rely on spatial information in terms of the RETFs $\H_T(l)$, cf. (\ref{eq:g}) and (\ref{eq:B}),
which needs to be estimated in practice.
The estimate $\hat{\H}_T(l)$ commonly contains estimation errors, i.e. we have $\hat{\H}_T(l) \neq \H_T(l)$.
Further, the RETF-based data model in (\ref{eq:sm:sumxne_matmult}) itself may be erroneous, e.g., due to dependencies across frequency bins \cite{avergel07}.
Finally, (\ref{eq:xell_xl_uncorrelated}) may be violated, e.g., due to overlapping windows in the STFT processing.
In general, these estimation and modeling errors cause incomplete blocking and therefore target component leakage through the BM, such that $\Exp [\u_\textsl{\tiny{SC}}(l) s_{T}^*(l)] \neq \mathbf{0}$, cf. (\ref{eq:q}), (\ref{eq:uSC}). 
This may be referred to as \textit{spatial target component leakage}.
Similarly, if $s_{T}(l)$ is temporally correlated such that (\ref{eq:xe_uncorrelated}) is violated, e.g., 
due to overlapping windows in the STFT processing or to too small lengths and window shifts, 
we find $\Exp [\u_\textsl{\tiny{LP}}(l) s_{T}^*(l)] \neq \mathbf{0}$, cf.  (\ref{eq:q}), (\ref{eq:uLP}), which may be referred to as \textit{temporal target component leakage}.

Potentially, spatial and temporal leakage cause a biased \cite{dietzen19TASLP} filter estimate $\hat{\w}(l)$, which leads to partial suppression of $s_{T}(l)$, also referred to as speech cancellation in GSC terminology \cite{doclo2010acoustic, gannot17}, or excessive whitening
in MCLP terminology \cite{delcroix07b}.
However, note that the Kalman filter offers inherent robustness towards target-component leakage.
To see this, consider the measurement update terms in (\ref{muw})--(\ref{muP}), respectively given by $\k(l)e^*(l)$ and $\k(l)\u^\herm(l)\bPsi_{\tilde{w}}(l)$.
Using (\ref{eq:obsEq}) and (\ref{eq:wtilde}), we may express $e^*(l)$ in (\ref{E}) in terms of $s^*_T(l)$, while using (\ref{Psie}), we may similarly express $\k(l)$ in (\ref{K}) in terms of ${\varphi}_{s_T}(l)$, i.e.
\begin{align}
e^*(l) &= s^*_T(l) -\u^\herm(l)\tilde{\w}(l),\label{eq:e_alt}\\
\k(l) &= \dfrac{\bPsi_{\tilde{w}}(l)\u(l)}{\u^\herm(l)\bPsi_{\tilde{w}}(l)\u(l) + {\varphi}_{s_T}(l)}\label{eq:k_alt}.
\end{align}
From (\ref{eq:e_alt})--(\ref{eq:k_alt}), we note that ${\varphi}_{s_T}(l) = \Exp [s_T(l)s^*_T(l)]$ acts as a regularization parameter in both update terms $\k(l)e^*(l)$ and $\k(l)\u^\herm(l)\bPsi_{\tilde{w}}(l)$.
Consequently, strong target powers inhibit the measurement update, while weak target powers promote it.
Put differently, in terms of robustness towards  target-component leakage and convergence, the Kalman filter benefits from non-stationarities and sparsity in ${\varphi}_{s_T}(l)$ across time. 
Note that in recursive MCLP implementations  based on the weighted prediction error (WPE) criterion and RLS  \cite{yoshioka09, yoshioka13, jukic2017, heymann18}, 
the target-component PSD similarly appears as a regularization term in the update equations.

In practice, we rely on estimates $\hat{\varphi}_{s_T}(l)$, which should hence maintain non-stationarities.
In WPE RLS literature, the target-component PSD estimate is obtained, e.g., directly from the plain microphone signals \cite{yoshioka09}, based on a late-reverberant PSD estimate obtained by means of an exponential decay model  \cite{yoshioka13, jukic2017}, or using a neural network \cite{heymann18}.
Here, as we consider a more generic signal model comprising several reverberant speech components and diffuse noise, cf. Sec. \ref{sec:signal}, we instead estimate $\hat{\varphi}_{s_T}(l)$ by means of \cite{dietzen19TBA}, cf. Sec. \ref{sec:targetPSDRETFupdate}.

\subsection{Target PSD Estimation and RETF Update}
\label{sec:targetPSDRETFupdate}

We require an RETF estimate $\hat{\H}_T(l)$ of $\H_T(l)$, cf. (\ref{eq:g}) and (\ref{eq:B}), and a PSD estimate $\hat{\varphi}_{s_{T}}(l)$ of $\varphi_{s_{T}}(l)$, cf. (\ref{Psie}) and (\ref{alt_gain}).
To this end, we use an algorithm recently proposed in \cite{dietzen19TBA} by the authors of this paper,  
which computes early PSD estimates and recursively updates the RETF estimates for all $N$ point sources.
The algorithm \cite{dietzen19TBA} is summarized as follows.

Let ${\bPsi}_{x_\textsl{e}}(l) = \Exp[\x_\textsl{e}(l)\x^\herm_\textsl{e}(l)]$ denote the correlation matrix of $\x_\textsl{e}(l)$ within frame $l$, which generally has rank $N$ and is given by
\begin{align}
{\bPsi}_{x_\textsl{e}}(l)  &= {\H}(l)\Diag[ {\bphi}_{s}(l)]{\H}^\herm(l),
\label{eq:sm:Psixe}\\
 {\bphi}_{s}(l) &=
 \begin{pmatrix}
 \varphi_{s_{1}}(l) & \cdots & \varphi_{s_{N}}(l)
 \end{pmatrix}^{\transp},
 \label{eq:sm:phixe}
  \end{align}
with $\varphi_{s_{n}}(l)$ denoting the PSD of the early speech source image $s_n(l)$.
Instead of directly using the conventional early correlation matrix model in (\ref{eq:sm:Psixe}), the algorithm in  \cite{dietzen19TBA} is based on its factorization, i.e. it relies on the square-root model
\begin{align}
{\bPsi}{}^{\nicefrac{1}{2}}_{x_\textsl{e}}(l)\bOmega(l) = \H(l)\Diag[\bphi^{\nicefrac{1}{2}}(l)], \label{eq:factorizedmodel}
\end{align}
where ${\bPsi}{}^{\nicefrac{1}{2}}_{x_\textsl{e}}(l) \in \mathbb{C}^{M\times N}$ and $\bphi^{\nicefrac{1}{2}} \in \mathbb{C}^N$
are some square roots of ${\bPsi}{}_{x_\textsl{e}}(l)$ and $ {\bphi}_{s}(l)$
such that ${\bPsi}{}^{\nicefrac{1}{2}}_{x_\textsl{e}}(l){\bPsi}^{\nicefrac{\herm}{2}}_{x_\textsl{e}}(l) = {\bPsi}_{x_\textsl{e}}(l)$ and $\Diag[\bphi^{\nicefrac{\herm}{2}}(l)]\bphi^{\nicefrac{1}{2}}(l) =  {\bphi}_{s}(l)$, respectively, and $\bOmega(l)$ is a unitary matrix, i.e. $\bOmega(l)\bOmega^\herm(l) = \mathbf{I}$, which accounts for the non-uniqueness of both square-roots. 
Note that right-multiplying each side of (\ref{eq:factorizedmodel}) with its Hermitian yields (\ref{eq:sm:Psixe}). 
In the estimation, we distinguish the prior and posterior RETF estimates $\hat{\H}(l)$ and $\hat{\H}^{+}(l)$, respectively, and assume that initial RETF estimates $\hat{\H}(0)$  are available, which may be based on, e.g., initial single-source RETF estimates acquired from segments with distinctly active sources \cite{markovichAug09}, or some initial knowledge or estimates of the associated dicrections of arrival (DoAs) \cite{scheuing2008correlation, chen2010introduction}.
Given a (to be obtained) square-root estimate $\hat{\bPsi}{}^{\nicefrac{1}{2}}_{x_\textsl{e}}(l)$ and a prior RETF estimate $\hat{\H}(l)$, which is propagated from the previous posterior, i.e. $\hat{\H}(l) = \hat{\H}^{+}(l\-1)$, 
we first obtain the unitary and diagonal estimates $\hat{\bOmega}(l)$ and $\Diag[\hat{\bphi}{}^{\nicefrac{1}{2}}]$, yielding $\hat{\bphi}_{s}(l) = \Diag[\hat{\bphi}{}^{\nicefrac{\herm}{2}}]\hat{\bphi}{}^{\nicefrac{1}{2}}$, and based on these estimates second update the RETF estimate, yielding the posterior $\hat{\H}^{+}(l)$, whereat the recursion is closed.
Here, both steps are based on approximation error minimization with respect to the square-root model in (\ref{eq:factorizedmodel}).
Given $\hat{\bphi}_{s}(l)$ and $\hat{\H}^{+}(l)$, we extract $\hat{\varphi}_{s_{T}}(l)$ and $\hat{\H}_T(l)$ as $\hat{\varphi}_{s_{T}}(l) = \mathbf{1}^\transp[\hat{\bphi}_{s}(l)]_{\in T}$ and $\hat{\H}_T(l) = [\hat{\H}^{+}(l)]_{\in T}$, cf. Sec. \ref{sec:signal}.

The said required square root ${\bPsi}{}^{\nicefrac{1}{2}}_{x_\textsl{e}}(l)$ is estimated in the following manner. 
While $\x_{n|\ell}(l)$ and $\v(l)$ exhibit a fundamentally different temporal behavior across frames, cf. Sec. \ref{sec:signal}, we assume that their spatial behavior within frame $l$ is the same.
Specifically, we model both ${\x}_{n|\ell}(l)$ and $\mathbf{v}(l)$ as spatially diffuse with coherence matrix $\mathbf{\Gamma}  \in \mathbb{C}^{M \times M}$, which may be computed from the microphone array geometry \cite{jacobsen2000coherence, DalDeganP18} and is therefore assumed to be known.
For the late reverberant component ${\x}_{n|\ell}(l)$, this is a commonly made assumption \cite{jacobsen2000coherence, Braun18TASLP, KodrasiD18}. 
For the noise component $\mathbf{v}(l)$, the assumption is commonly made for noise types such as, e.g., babble noise \cite{habets2008generating}, which we use in our simulations, cf. Sec. \ref{sec:sim}.
Based on these assumptions, the microphone signal correlation matrix ${\bPsi}_{y}(l) = \Exp[\y(l)\y^\herm(l)]$ may be written as
\begin{align}
{\bPsi}_{y}(l)  = {\bPsi}_{x_\textsl{e}}(l) + {\varphi}_{\textsl{d}}(l)\mathbf{\Gamma}, 
\label{eq:sm:Psiy}
\end{align}
with ${\varphi}_{\textsl{d}}(l) = \sum_{n=1}^N \varphi_{x_{n|\ell}}(l) +  {\varphi}_v(l)$ and $\varphi_{x_{n|\ell}}(l)$ and ${\varphi}_v(l)$ denoting the PSDs of the  late-reverberant speech components and the diffuse noise component, respectively.
We obtain a subspace representation of (\ref{eq:sm:Psiy}) by means of the generalized eigenvalue decomposition (GEVD) of ${\bPsi}_{y}(l)$ and $\mathbf{\Gamma}$. 
Based on the generalized eigenvectors and generalized eigenvalues, ${{\bPsi}}_{y}(l)$ may be decomposed into a diffuse component, cf. also the diffuse PSD estimator in \cite{KodrasiD18}, and a factorized early rank-$N$ component ${\bPsi}_{x_\textsl{e}}(l) = {\bPsi}{}^{\nicefrac{1}{2}}_{x_\textsl{e}}(l){\bPsi}^{\nicefrac{\herm}{2}}_{x_\textsl{e}}(l)$.
A temporally smooth estimate ${\hat{\bPsi}}_{y|\textsl{sm}}(l)$ of ${{\bPsi}}_{y}(l)$ itself is obtained from the microphone signals by recursively averaging $\y^\herm(l)\y(l)$.
In order to restore non-stationarities, we desmooth\footnote{Considering recursive averaging as an invertible recursive filtering operation, the generalized eigenvalues may be desmoothed by means of the corresponding inverse filter.} the generalized eigenvalues of $\hat{{\bPsi}}_{y|\textsl{sm}}(l)$ and $\mathbf{\Gamma}$ and thereby yield non-stationary PSD estimates in the subsequent processing steps, as as favored in the Kalman filter, cf. Sec \ref{sec:leakage}.
For further details, we refer the interested reader to \cite{dietzen19TBA}.

Considering recursive averaging as an invertible recursive filtering operation, the generalized eigenvalues may be desmoothed by means of the corresponding inverse filter.

\subsection{Process Equation Parameter Tuning and Initialization}
\label{sec:initproc}

The tracking and convergence behavior of the Kalman filter depends on its process equation parameter tuning and initialization.
The process equation models the evolution of the state by means of the parameters  $\A(l)$ and $\bPsi_{w_{\hspace{-.05em}\scriptscriptstyle{\Delta}}}(l)$, cf. (\ref{eq:procEq}) and (\ref{tuw})--(\ref{tuP}).
In practice, only limited knowledge of the state evolution is available, such that $\A(l)$ and $\bPsi_{w_{\hspace{-.05em}\scriptscriptstyle{\Delta}}}(l)$ are commonly left  to tuning \cite{enzner06, braun16, dietzen17, BraunJune2018}.
Typically, both $\A(l)$ and $\bPsi_{w_{\hspace{-.05em}\scriptscriptstyle{\Delta}}}(l)$ are chosen to be scaled identities, 
with $\A(l)$ commonly time-invariant \cite{enzner06, braun16, dietzen17, BraunJune2018} and acting as a forgetting factor \cite{enzner06, dietzen17},
and $\bPsi_{w_{\hspace{-.05em}\scriptscriptstyle{\Delta}}}(l)$ either time-variant \cite{enzner06, braun16, BraunJune2018} or time-invariant \cite{dietzen17}.
Here, we set $\A(l)$ and $\bPsi_{w_{\hspace{-.05em}\scriptscriptstyle{\Delta}}}(l)$ based on the assumption that the state correlation matrix
$\bPsi_{w}(l)$ is time-invariant, i.e. $\bPsi_{w}(l) = \bPsi_{w}$.
Unfortunately, $\bPsi_{w}$ is unknown and not available in practice, however, we may define a rough guess $\bar{\bPsi}_{w}$. 
Given such a guess $\bar{\bPsi}_{w}$, by means of a forgetting factor $\alpha \in (0, 1)$, we may account for a steadily time-varying acoustic scenario and true state $\w(l)$ by setting
\begin{align}
\A(l) &= \sqrt{\alpha}\,\mathbf{I}, \label{eq:Atuned}\\
\bPsi_{w_{\hspace{-.04em}\scriptscriptstyle{\Delta}}}(l) &= (1-\alpha)\bar{\bPsi}_{w},\label{eq:Psiwdelta}
\end{align}
such that if $\bar{\bPsi}_{w} = {\bPsi}_{w}$, we rightly have ${\bPsi}_{w} = \alpha {\bPsi}_{w} + (1-\alpha){\bPsi}_{w}$ from (\ref{eq:procEq}).
While $\bar{\bPsi}_{w}$ may rather be defined by design than by truly estimating $\bPsi_{w}$,
the notion of $\bar{\bPsi}_{w}$ being a rough guess of $\bPsi_{w}$ may nonetheless guide its definition to some extent. 
Here, we choose a diagonal matrix with distinct diagonal elements.
With $\bar{\bPsi}{}_{{w}} = \Diag[\bar{\bpsi}{}_{{w}}]$, 
let $\bar{\bpsi}{}_{{w}_{\textsl{\tiny{SC}}}} \in \mathbb{R}^{M-N_T}$ and $\bar{\bpsi}{}_{{w}_{\textsl{\tiny{LP}}}}  \in \mathbb{R}^{(L-1)M}$ denote the subvectors of $\bar{\bpsi}{}_{{w}}$ associated to the SC and the LP filter, respectively, which we treat separately.
Expecting lower values for later prediction coefficients in the LP filter, we choose the power of the diagonal elements in $\bar{\bpsi}{}_{{w}_{\textsl{\tiny{LP}}}}$ to drop exponentially each $M$ elements, i.e. we set
\begin{align}
\bar{\bpsi}{}_{{w}_{\textsl{\tiny{SC}}}} &= \bar{\psi}{}_{{w}_{\textsl{\tiny{SC}}}}\mathbf{1},\label{eq:psiwSC}\\
\bar{\bpsi}{}_{{w}_{\textsl{\tiny{LP}}}} &= \begin{pmatrix}
 \bar{\psi}_{{w}_{\textsl{\tiny{LP}}}}^1 \mathbf{1}^\transp & \dots &  \bar{\psi}_{{w}_{\textsl{\tiny{LP}}}}^{L-1} \mathbf{1}^\transp
\end{pmatrix}^\transp, \label{eq:psiwLP}
\end{align}
with $ \bar{\psi}_{{w}_{\textsl{\tiny{SC}}}} > 0$ and $ \bar{\psi}_{{w}_{\textsl{\tiny{LP}}}} \in (0, 1)$ further adjustable.

The matrix $\bar{\bPsi}_{w}$ may also be used to initialize the Kalman filter.
With the commonly chosen initial state estimate $\hat{\w}(0) = \mathbf{0}$, we have $\tilde{\w}(0) = {\w}(0)$ in (\ref{eq:wtilde}), such that the true initial state estimation error correlation matrix $\bPsi_{\tilde{w}}(0)$ becomes $\bPsi_{\tilde{w}}(0) = \bPsi_{{w}}(0) = \bPsi_{{w}}$.
Therefore, we initialize the Kalman filter by 
\begin{align}
\hat{\w}(0) &= \mathbf{0},  \\
\hat{\bPsi}_{\tilde{w}}(0) &= \bar{\bPsi}_{w}, \label{eq:Pinit}
\end{align}
in (\ref{tuw})--(\ref{tuP}), where $\hat{\bPsi}_{\tilde{w}}(0)$ in (\ref{eq:Pinit}) is an estimate if $\bar{\bPsi}_{w} \neq \bPsi_{{w}}$.
Finally, note that the process equation parameter tuning in (\ref{eq:Atuned})--(\ref{eq:Psiwdelta}) may also be considered from a (re-)initialization perspective.
In case of meaningful measurement updates, the Kalman filter tracks ${\w}(l)$, but otherwise tends to return to its initial condition due to (\ref{eq:Atuned})--(\ref{eq:Psiwdelta}), such that explicit re-initialization as, e.g., in case of a sudden change in the acoustic environment, is not necessary.
To see this, consider the case where, e.g.,  $\u(l) = \mathbf{0}$ for a period of time, such that no measurement update is performed. 
In this case, regardless of their current values, we have $\hat{\w}(l)$ slowly converging to $\mathbf{0}$ and $\hat{\bPsi}_{\tilde{w}}(l)$ slowly converging to $\bar{\bPsi}{}_{{w}}$, cf. (\ref{tuw})--(\ref{tuP}).
Note that if desired, explicit re-initialization may still easily be incorporated in the proposed concept, namely by defining $\alpha$ time-variant and setting it to zero at the determined re-initialization point. 

\section{Simulations}
\label{sec:sim}

In order to demonstrate the effectiveness of the presented ISCLP Kalman filter, we define two case studies, case A and case B.
In case A, we compare to the (computationally more demanding) alternating Kalman filters proposed in \cite{BraunJune2018}.
Here, we consider one reverberant speech and a babble noise component, ${\x}_1(l)$ and ${\v}(l)$, with ${\x}_1(l)$ containing the target component $\x_{\textsl{e}|T}(l) = \x_{1|\textsl{e}}(l)$.
In case B, we compare to a  (computationally more demanding) MCLP+GSC Kalman filter cascade, which conceptually relates to \cite{delcroix15, yoshioka15} in that it cascades linear prediction and beamforming.
Here, we consider two reverberant speech components and a babble noise component, ${\x}_1(l)$, ${\x}_2(l)$, and ${\v}(l)$, with ${\x}_1(l)$ again containing the target component $\x_{\textsl{e}|T}(l) = \x_{1|\textsl{e}}(l)$, and ${\x}_2(l)$ an interfering speech component to be canceled.
In both cases, we investigate the algorithms' behavior depending on the signal-to-noise ratio, $\mathit{SNR}$, which is defined as the power ratio of ${\x}_1(l)$ to ${\v}(l)$, and depending on the filter length $L$.
In case A, we additionally investigate the convergence behavior.

In what follows, we describe the two reference algorithms in more detail in Sec. \ref{sec:refalg}, the performance measures in Sec. \ref{sec:perfmeas}, the acoustic scenario in Sec. \ref{sec:acoustscen}, the algorithmic settings in \ref{sec:algoset}, and finally the simulation results in Sec. \ref{sec:results}.
 
\subsection{Reference Algorithms}
\label{sec:refalg} 

We discuss the alternating Kalman filters  in Sec. \ref{sec:A_altKF} and the MCLP+GSC Kalman filter cascade in Sec. \ref{sec:B_MCLP+GSC}.

\subsubsection{Case A: Alternating Kalman Filters}
\label{sec:A_altKF}

In \cite{BraunJune2018}, MCLP-based dereverberation and noise reduction is performed in each microphone channel using two alternating Kalman filters.
The Kalman filter dedicated to dereverberation estimates a multiple-output LP filter, and the Kalman filter dedicated to noise reduction estimates the noise-free reverberant speech component. 
The enhanced signal is computed from the posterior state estimates of both Kalman filters.
The two state vectors have dimensions $M^2(L-1)$ and $M(L-1)$, respectively, while the ISCLP Kalman filter requires a single state vector with dimension $ML-N_T$ only with $N_T=1$ in case A, cf. (\ref{eq:wstacked}). 
Since the Kalman filter in general exhibits a quadratic computational cost in the state vector dimension, 
the alternating Kalman filters are computationally roughly $M^2$ times as demanding as the ISCLP Kalman filter.
The two state space models do not provide a spatial distinction between point sources (and therefore do not require RETF estimates, as opposed to the ISCLP Kalman filter) and further do not consider temporally correlated interference components such as interfering reverberant speech. We hence set ${\x}_2(l) = \mathbf{0}$ when comparing to \cite{BraunJune2018}, i.e. interfering speech is absent, cf. Sec. \ref{sec:caseAwith}.

The alternating Kalman filters require correlation matrix estimates of the measurement and process noises, more precisely of the random variation of the multiple-output LP filter state, comparable to $\bPsi_{w_{\hspace{-.05em}\scriptscriptstyle{\Delta}}}(l)$ in the ISCLP Kalman filter, cf. (\ref{eq:procEq}), the early component ${\bPsi}_{x_\textsl{e}|T}(l) = {\bPsi}_{x_\textsl{e}}(l)$, the early-plus-noise component $ {\bPsi}_{x_\textsl{e}}(l) + {\bPsi}_{v}(l)$, and the noise component ${\bPsi}_{v}(l)$  \cite{BraunJune2018}.
In the original implementation in \cite{BraunJune2018}, a time-invariant estimate $\hat{\bPsi}_{v}$ is assumed to be available, which we here compute in an oracle fashion from $\v(l)$ directly, 
while the other correlation matrices are estimated based on the previous state estimates and error signals of the alternating Kalman filters. 
For the sake of a fair and more meaningful comparison, we implement two versions of \cite{BraunJune2018}. 
The first version is implemented as proposed in \cite{BraunJune2018} and discussed above, subsequently referred to as the original alternating Kalman filters. 
In the second version, we align the parameter estimation and tuning towards the proposed approach, i.e.  
${\bPsi}_{x_\textsl{e}}(l)$ is instead estimated based on \cite{dietzen19TBA}, cf. Sec. \ref{sec:targetPSDRETFupdate}, 
and the process equation parameters modeling the evolution of the multiple-output LP filter state are defined similarly to Sec. \ref{sec:initproc}, subsequently referred to as the modified alternating Kalman filters.

\subsubsection{Case B: MCLP+GSC Kalman Filter Cascade}
\label{sec:B_MCLP+GSC}

In \cite{delcroix15, yoshioka15}, multiple-output MCLP based on the (iterative) WPE criterion \cite{yoshioka11, yoshioka12} is cascaded with MVDR beamforming in order to reduce noise after dereverberation, which became a popular approach in the CHiME-5 challenge \cite{chime5}.
For the sake of a close comparison, however, we here instead compare to a (recursive) multiple-output MCLP-based Kalman filter cascaded with a (recursive) GSC-based Kalman filter, subsequently referred to as MCLP+GSC.
Herein, we estimate the LP and SC filters independently.
The enhanced signal at the GSC output is computed using spectral post-processing of the same kind as in (\ref{alt_gain})--(\ref{e_post_altGain}).
The two state vectors have dimensions $M^2(L-1)$ and $M-1$, respectively, while the ISCLP Kalman filter requires a single state vector with dimension $ML-N_T$ only with $N_T=1$ in case B, cf. (\ref{eq:wstacked}). 
Since the Kalman filter in general exhibits a quadratic computational cost in the state vector dimension, 
the MCLP+GSC Kalman filter cascade is computationally roughly $M^2$ times as demanding as the ISCLP Kalman filter.
The GSC state space model does provide a spatial distinction between point sources (based on an RETF estimate, as the ISCLP Kalman filter). 
We hence set ${\x}_2(l) \neq \mathbf{0}$ when comparing to the MCLP+GSC Kalman filter cascade,  i.e. interfering speech is present, cf. Sec. \ref{sec:caseAwith}.

The MCLP and GSC Kalman filters require correlation matrix estimates of their respective measurement and process noises, more precisely of the random variation of the multiple-output LP filter and SC filter state, respectively, defined similarly to the corresponding SC and LP submatrices of $\bPsi_{w_{\hspace{-.05em}\scriptscriptstyle{\Delta}}}(l)$ in the ISCLP Kalman filter, cf. (\ref{eq:procEq}), and the early components ${\bPsi}_{x_\textsl{e}|T}(l)$ and ${\varphi}_{s_T}(l)$, respectively, computed based on \cite{dietzen19TBA} as in the proposed ISCLP Kalman filter, cf. Sec. \ref{sec:targetPSDRETFupdate}.

 \setlength\fwidth{16cm}
 \begin{figure*}[t]
\centering
\hspace*{-0.205cm}
    \setlength\fheight{8.6cm} 
%
%
\begin{tikzpicture}

\begin{axis}[%
width=\fwidth,
height=0.021\fheight,
at={(0\fwidth,0.499\fheight)},
scale only axis,
point meta min=-55,
point meta max=5,
axis on top,
xmin=0.5,
xmax=61.5,
xtick={6,16,26,36,46,56},
xticklabels={{$-50$},{$-40$},{$-30$},{$-20$},{$-10$},{$0$}},
separate axis lines,
every outer y axis line/.append style={black},
every y tick label/.append style={font=\color{black}},
ymin=0.5,
ymax=2.5,
ytick={1},
yticklabels={{\rotatebox{-90}{$/$\si{dB}}}},
axis background/.style={fill=white},
yticklabel pos=right,
major tick length = 0em, ylabel style={at={(axis description cs:0.045,0.5)}, xshift=.0em, anchor=north}, yticklabel style={rotate=90}, xlabel style={at={(ticklabel cs: -0.01,-8)}, anchor=west}, title style={at={(axis description cs:.92,0)}, xshift=0.2em, yshift=-1.1em, anchor=south, font=\normalfont}
]
\addplot [forget plot] graphics [xmin=0.5,xmax=61.5,ymin=0.5,ymax=2.5] {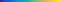};
\end{axis}

\begin{axis}[%
width=0.233\fwidth,
height=0.433\fheight,
at={(0\fwidth,0\fheight)},
scale only axis,
point meta min=-55,
point meta max=5,
axis on top,
xmin=0.5,
xmax=128.5,
xtick={1,64},
xticklabels={{$0$},{$1$}},
xlabel={\small{$t/$\si{s}}},
ymin=0.5,
ymax=257.5,
ytick={33,97,161,225},
yticklabels={{$1$},{$3$},{$5$},{$7$}},
ylabel={\small{$f/$\si{kHz}}},
axis background/.style={fill=white},
title style={font=\bfseries},
title={\colorbox{white}{(a)}},
major tick length = 0em, ylabel style={at={(axis description cs:0.045,0.5)}, xshift=.0em, anchor=north}, yticklabel style={rotate=90}, xlabel style={at={(ticklabel cs: -0.01,-8)}, anchor=west}, title style={at={(axis description cs:.92,0)}, xshift=0.2em, yshift=-1.1em, anchor=south, font=\normalfont}
]
\addplot [forget plot] graphics [xmin=0.5,xmax=128.5,ymin=0.5,ymax=257.5] {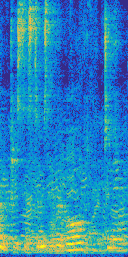};
\end{axis}

\begin{axis}[%
width=0.233\fwidth,
height=0.433\fheight,
at={(0.256\fwidth,0\fheight)},
scale only axis,
point meta min=-55,
point meta max=5,
axis on top,
xmin=0.5,
xmax=128.5,
xtick={1,64},
xticklabels={{$0$},{$1$}},
xlabel={\small{$t/$\si{s}}},
ymin=0.5,
ymax=257.5,
ytick={\empty},
axis background/.style={fill=white},
title style={font=\bfseries},
title={\colorbox{white}{(b)}},
major tick length = 0em, ylabel style={at={(axis description cs:0.045,0.5)}, xshift=.0em, anchor=north}, yticklabel style={rotate=90}, xlabel style={at={(ticklabel cs: -0.01,-8)}, anchor=west}, title style={at={(axis description cs:.92,0)}, xshift=0.2em, yshift=-1.1em, anchor=south, font=\normalfont}
]
\addplot [forget plot] graphics [xmin=0.5,xmax=128.5,ymin=0.5,ymax=257.5] {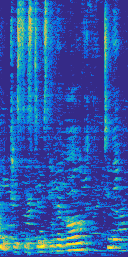};
\end{axis}

\begin{axis}[%
width=0.233\fwidth,
height=0.433\fheight,
at={(0.511\fwidth,0\fheight)},
scale only axis,
point meta min=-55,
point meta max=5,
axis on top,
xmin=0.5,
xmax=128.5,
xtick={1,64},
xticklabels={{$0$},{$1$}},
xlabel={\small{$t/$\si{s}}},
ymin=0.5,
ymax=257.5,
ytick={\empty},
axis background/.style={fill=white},
title style={font=\bfseries},
title={\colorbox{white}{(c)}},
major tick length = 0em, ylabel style={at={(axis description cs:0.045,0.5)}, xshift=.0em, anchor=north}, yticklabel style={rotate=90}, xlabel style={at={(ticklabel cs: -0.01,-8)}, anchor=west}, title style={at={(axis description cs:.92,0)}, xshift=0.2em, yshift=-1.1em, anchor=south, font=\normalfont}
]
\addplot [forget plot] graphics [xmin=0.5,xmax=128.5,ymin=0.5,ymax=257.5] {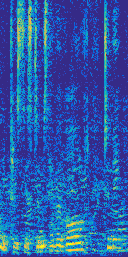};
\end{axis}

\begin{axis}[%
width=0.233\fwidth,
height=0.433\fheight,
at={(0.767\fwidth,0\fheight)},
scale only axis,
point meta min=-55,
point meta max=5,
axis on top,
xmin=0.5,
xmax=128.5,
xtick={1,64},
xticklabels={{$0$},{$1$}},
xlabel={\small{$t/$\si{s}}},
ymin=0.5,
ymax=257.5,
ytick={\empty},
axis background/.style={fill=white},
title style={font=\bfseries},
title={\colorbox{white}{(d)}},
major tick length = 0em, ylabel style={at={(axis description cs:0.045,0.5)}, xshift=.0em, anchor=north}, yticklabel style={rotate=90}, xlabel style={at={(ticklabel cs: -0.01,-8)}, anchor=west}, title style={at={(axis description cs:.92,0)}, xshift=0.2em, yshift=-1.1em, anchor=south, font=\normalfont}
]
\addplot [forget plot] graphics [xmin=0.5,xmax=128.5,ymin=0.5,ymax=257.5] {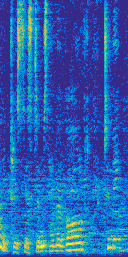};
\end{axis}
\end{tikzpicture}%
\caption{
Exemplary spectrograms depicting $2\,$\si{s} of (a) the reference microphone signal $y_1(l)$, and the corresponding outputs of (b) the original alternating Kalman filters, (c) the modified alternating Kalman filters, and (d) the ISCLP Kalman filter for $L=6$ at $\mathit{SNR} = 10\,$\si{dB}.
}
\label{fig:spectrograms}
\end{figure*}

\subsection{Performance Measures}
\label{sec:perfmeas}

As performance measures, we choose the perceptual evaluation of speech quality \cite{itu01}, $\mathit{PESQ}$, with mean opinion scores of objective listening quality $\in [1,4.5]$, 
the short-time objective intelligibility \cite{taal11},  $\mathit{STOI}$, with scores $\in [0,1]$, 
the frequency-weighted segmental signal-to-interference ratio \cite{loizou2007speech, hu2008evaluation
}, $\mathit{SIR}^{\mathit{fws}}$, in \si{dB}, 
and the cepstral distance \cite{loizou2007speech, hu2008evaluation
}, $\mathit{CD}$, in \si{dB}.
While high values are preferable for $\mathit{PESQ}$,  $\mathit{STOI}$, and  $\mathit{SIR}^{\mathit{fws}}$, low values are preferred for $\mathit{CD}$.
These intrusive measures require a clean reference signal $\tilde{s}_{T}(l)$, which approximates the target signal ${s}_{T}(l)$ in (\ref{eq:sm:st}).
In order to generate $\tilde{s}_{T}(l)$, we convolve the target speech source signal with the early part of the RIR to the first microphone, cf. Sec \ref{sec:acoustscen}, whereat we define the first $N_{\textsl{STFT}}$ samples of the RIR as its early part, with $N_{\textsl{STFT}}$ the analysis and synthesis window length of the STFT processing corresponding to $32\,\si{ms}$, cf. Sec. \ref{sec:algoset}.
Note that due to modeling errors in the RETF-model in (\ref{eq:sm:sumxne_matmult}), we generally have $\tilde{s}_{T}(l) \neq {s}_{T}(l)$.
When investigating the dependency on  $\mathit{SNR}$ or $L$, we compute the measures  from $4\,\si{s}$ to $10\,\si{s}$, i.e. roughly after convergence.
When investigating the convergence behavior, we compute the measures within sliding windows of $2\,\si{s}$ each.
The computed measures are averaged over several individual simulations, cf. Sec. \ref{sec:acoustscen}. 

\subsection{Acoustic Scenario}
\label{sec:acoustscen}

We describe the acoustic scenarios without and with interfering speaker in Sec. \ref{sec:caseAwithout} and Sec. \ref{sec:caseAwith}, respectively.

\subsubsection{Case A: Without Interfering Speech}
\label{sec:caseAwithout}
In case A, the microphone signals are composed of one reverberant speech and a babble noise component, ${\x}_1(l)$ and ${\v}(l)$, with ${\x}_1(l)$ containing the target component $\x_{\textsl{e}|T}(l) = \x_{1|\textsl{e}}(l)$.
To generate $\x_1(l)$, we use measured RIRs of $0.61\,$\si{s} reverberation time to a linear microphone array with $M = 5$ microphones and $8\,$\si{cm} inter-microphone distance \cite{madivae}.
When investigating the dependency on $\mathit{SNR}$ or $L$, the speech source remains positioned in $2\,$\si{m} distance of the microphone array at $0^\circ$ relative to the broad-side direction during $10\,$\si{s} of simulation.
When investigating the convergence behavior, the speech source remains positioned in $2\,$\si{m} distance at $0^\circ$ for the first $8\,$\si{s}, then jumping to $15^\circ$, where it remains for another $10\,$\si{s}.
Both female and male speech \cite{bando92} are used as speech source signals. 
The babble noise component is generated using \cite{habets2008generating, Audiotec}. 
From the speech source signal files and the babble noise file \cite{Audiotec}, we randomly select individual segments, yielding  individual simulations to be averaged in the performance evaluation, cf. Sec. \ref{sec:perfmeas}.
In total, when investigating the dependency on  $\mathit{SNR}$ or $L$, we generate $64$ individual simulations per condition.  
When investigating the convergence behavior, we generate $128$ individual simulations.

\subsubsection{Case B: With Interfering Speech}
\label{sec:caseAwith}

In case B, the microphone signals are composed of two reverberant speech components and a noise component, ${\x}_1(l)$, ${\x}_2(l)$, and ${\v}(l)$, with ${\x}_1(l)$ again containing the target component $\x_{\textsl{e}|T}(l) = \x_{1|\textsl{e}}(l)$, and ${\x}_2(l)$ an interfering speech component.
We investigate the dependency on $\mathit{SNR}$ and $L$, and generate ${\x}_1(l)$ and ${\v}(l)$ in the same manner as in case A, cf. Sec. \ref{sec:caseAwithout}.
To generate $\x_2(l)$, we use the same set of RIR measurements, where the associated source is positioned in $2\,$\si{m} distance at either $\{30,  60,  90\}^\circ$.
If ${\x}_1(l)$ contains female speech, then ${\x}_2(l)$ contains male speech \cite{bando92} and vice versa.
On average, $\x_1(l)$ and $\x_2(l)$ have roughly the same power.
From the speech source signal files and the babble noise file, we randomly select individual segments,  generating $3 \cdot 64 = 192$ individual simulations per condition to be averaged in the performance evaluation, cf. Sec. \ref{sec:perfmeas}.

\subsection{Algorithmic Settings}
\label{sec:algoset}

In our simulations, the sampling frequency is $f_s = 16\,$\si{kHz}, and the STFT analysis and synthesis uses square-root Hann windows of $N_{\textsl{STFT}} = 512$ samples with $50\%$ overlap.
When investigating the dependency on $\mathit{SNR}$ and the convergence behavior, we set $L = 6$ in (\ref{eq:uLP}).
The estimates $\hat{\bphi}_{s}(l)$ and $\hat{\H}^{+}(l)$, required in (\ref{Psie}) and (\ref{eq:g}), (\ref{eq:B}) are obtained by means of \cite{dietzen19TBA}, cf. Sec. \ref{sec:targetPSDRETFupdate}.
In (\ref{eq:Atuned})--(\ref{eq:Psiwdelta}), we set $\alpha$ such that $10 \log_{10}(1-\alpha) = -25 \si{dB}$.
Expecting lower values for SC filter coefficients at higher frequencies due to generally reduced spatial correlations between individual microphones, we choose  $\bar{\psi}_{{w}_{\textsl{\tiny{SC}}}}$ in (\ref{eq:psiwSC}) to be frequency-dependent with  $10 \log_{10}\bar{\psi}_{{w}_{\textsl{\tiny{SC}}}}$ decreasing linearly from $0\,\si{dB}$ at $0\,\si{kHz}$ to $-15\,\si{dB}$ at $8\,\si{kHz}$.
In (\ref{eq:psiwLP}), we set $10 \log_{10}\bar{\psi}_{{w}_{\textsl{\tiny{LP}}}} = -4\,\si{dB}$.
In (\ref{alt_gain}), we set $\beta$ such that $20 \log_{10}\beta = -2\,\si{dB}$, and $\gamma(0) = 1$.
 
\subsection{Results}
\label{sec:results}

We discuss the results in case A  and B in Sec. \ref{sec:rescaseA} and Sec. \ref{sec:rescaseB}, respectively.
Audio examples are available at \cite{taslp19bCodeAudio}.

\subsubsection{Case A}
\label{sec:rescaseA}

\setlength\fwidth{7.5cm}
\begin{figure}[t]
\centering
\hspace*{-0.218cm}
    \setlength\fheight{12cm} 
    \input{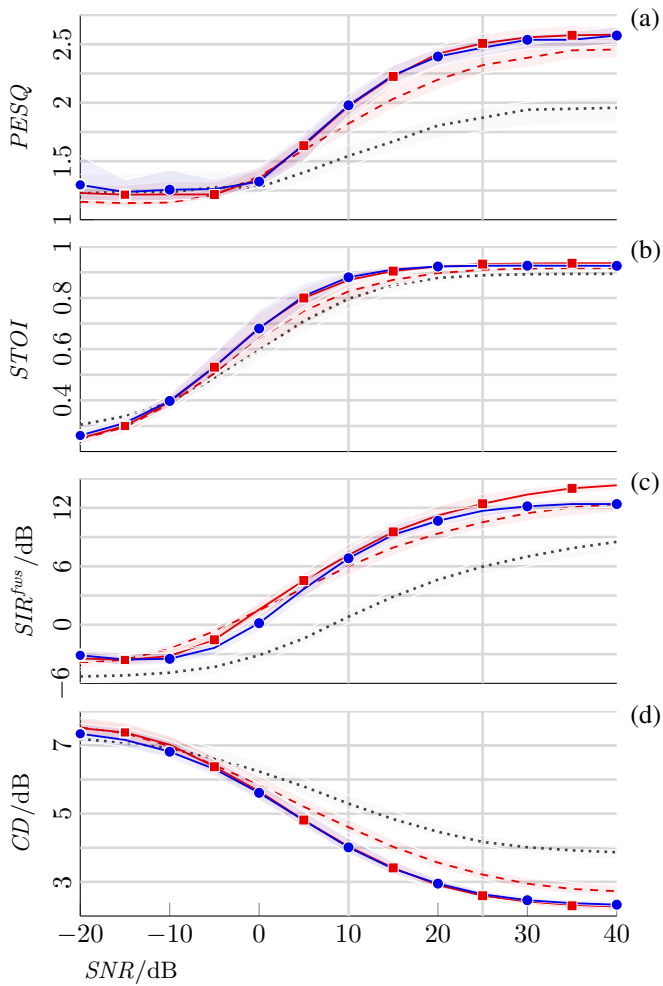} 
\caption{(a) $\mathit{PESQ}$, (b) $\mathit{STOI}$,  (c) $\mathit{SIR}^{fws}$, and (d) $\mathit{CD}$  versus $\mathit{SNR}$ for the reference microphone signal [\ref{TASLP_ISCLP_altKF_SNR_3}],  the original alternating Kalman filters [\ref{TASLP_ISCLP_altKF_SNR_6}], the modified alternating Kalman filters [\ref{TASLP_ISCLP_altKF_SNR_11}], and the ISCLP Kalman filter [\ref{TASLP_ISCLP_altKF_SNR_16}] with $L = 6$ if
interfering speech is absent.}
\label{fig:altKFSNR}
\end{figure}

\setlength\fwidth{7.14cm}
\begin{figure}[t]
\centering
\hspace*{-0.218cm}
    \setlength\fheight{12cm} 
    \input{./fig/TASLP_ISCLP_altKF_L_SNR25.tex} 
\caption{(a) $\Delta\mathit{PESQ}$, (b) $\Delta\mathit{STOI}$,  (c) $\Delta\mathit{SIR}^{fws}$, and (d) $\Delta\mathit{CD}$  versus $L$ with respect to the reference microphone signal for the original alternating Kalman filters [\ref{TASLP_ISCLP_altKF_SNR_6}], the modified alternating Kalman filters [\ref{TASLP_ISCLP_altKF_SNR_11}], and the ISCLP Kalman filter [\ref{TASLP_ISCLP_altKF_SNR_16}] at $\mathit{SNR} = 25\,$\si{dB} if
interfering speech is absent.
}
\label{fig:altKFL}
\end{figure}

\setlength\fwidth{7.14cm}
\begin{figure}[t]
\centering
\hspace*{-0.218cm}
    \setlength\fheight{12cm} 
    \input{./fig/TASLP_ISCLP_altKF_CONV.tex} 
\caption{(a) $\Delta\mathit{PESQ}$, (b) $\Delta\mathit{STOI}$,  (c) $\Delta\mathit{SIR}^{fws}$, and (d) $\Delta\mathit{CD}$  versus $t$ with respect to the reference microphone signal for the original alternating Kalman filters [\ref{TASLP_ISCLP_altKF_SNR_6}], the modified alternating Kalman filters [\ref{TASLP_ISCLP_altKF_SNR_11}], and the ISCLP Kalman filter [\ref{TASLP_ISCLP_altKF_SNR_16}] with $L = 6$ at $\mathit{SNR} = 10\,$\si{dB} if
interfering speech is absent.
}
\label{fig:altKFconv}
\end{figure}

Consider the spectrograms in Fig. \ref{fig:spectrograms} depicting $2\,$\si{s} of (a) the reference microphone signal $y_1(l)$, and the corresponding outputs of (b) the original alternating Kalman filters, (c) the modified alternating Kalman filters, and (d) the ISCLP Kalman filter for $L=6$ in an exemplary simulation at $\mathit{SNR} = 10\,$\si{dB}.
As can be seen by comparison with (a), all three algorithms in (b)--(d) considerably reduce reverberation and noise. 
Yet, their spectrograms exhibit slightly different features. 
As opposed to the modified alternating Kalman filters and the ISCLP Kalman filter (c)--(d), the original alternating Kalman filters (b) show some amount of temporal smearing resembling musical noise \cite{BraunJune2018}. 
This is due to errors in the correlation matrix estimates used to update the alternating Kalman filters, which in turn are computed recursively based on the alternating Kalman filters' previous state estimates and error signals \cite{BraunJune2018}. 
In contrast, in the modified alternating Kalman filters and the ISCLP Kalman filter, the required correlation matrix estimates and PSD estimates are computed directly from the microphone signals while maintaining non-stationarities, cf. Sec. \ref{sec:targetPSDRETFupdate} and Sec. \ref{sec:A_altKF}. 
As compared to the modified alternating Kalman filters (c), the signal power in the ISCLP Kalman filter (d) decays somewhat less quickly after transient speech components, which is due to $\beta > 0$ in  (\ref{alt_gain}), cf. Sec. \ref{sec:algoset}, resulting in a perceptually somewhat more pleasant sound image \cite{taslp19bCodeAudio}. 

Fig. \ref{fig:altKFSNR} shows the performance in terms of (a) $\mathit{PESQ}$, (b) $\mathit{STOI}$,  (c) $\mathit{SIR}^{fws}$, and (d) $\mathit{CD}$  versus $\mathit{SNR}$ for the reference microphone signal [\ref{TASLP_ISCLP_altKF_SNR_3}],  the original alternating Kalman filters [\ref{TASLP_ISCLP_altKF_SNR_6}], the modified alternating Kalman filters [\ref{TASLP_ISCLP_altKF_SNR_11}], and the ISCLP Kalman filter [\ref{TASLP_ISCLP_altKF_SNR_16}] with $L = 6$.
In this and the following figures, the graphs denote medians over all individual simulations,
cf. Sec. \ref{sec:acoustscen}, and the shaded areas indicate the range from the first
to the third quartile.
Overall, the measures show a high degree of agreement.  
As expected, the reference microphone signal reaches better scores at higher $\mathit{SNR}$ values  in all measures.
Above roughly $\mathit{SNR} = -5\,\si{dB}$, all three algorithms show a significant improvement over the reference microphone signal in all measures, least pronounced in $\mathit{STOI}$.
The modified alternating Kalman filters generally outperform the original alternating Kalman filters, validating the modified parameter estimation and tuning aligned to the proposed ISCLP Kalman filter, cf. Sec. \ref{sec:A_altKF}.
In terms of $\mathit{PESQ}$,  $\mathit{STOI}$, and $\mathit{CD}$, the ISCLP Kalman filter reaches very similar scores as compared to the modified alternating Kalman filters. 
In terms of $\mathit{SIR}^{fws}$, the ISCLP Kalman filter performs somewhat worse than the modified alternating Kalman filters above $\mathit{SNR} = 20\,\si{dB}$, which is due to a small amount of speech cancellation caused by the SC filter, cf. Sec. \ref{sec:leakage}. 
Note that in this $\mathit{SNR}$ range, the babble noise component $\v(l)$ becomes negligible, i.e. reverberant interference is pre-dominant, which can be handled by the LP filter only. 
The SC filter therefore becomes superfluous in this case.  
Further simulations showed that the ISCLP Kalman filter may reach similar $\mathit{SIR}^{fws}$ scores as compared to the modified alternating Kalman filters if the SC filter variance $\bar{\psi}_{{w}_{\textsl{\tiny{SC}}}}$ in (\ref{eq:psiwSC}) is set depdending on the $\mathit{SNR}$, which allows to essentially switch off the SC filter at high $\mathit{SNR}$ values, and thereby avoid unnecessary speech cancellation.

Fig. \ref{fig:altKFL} depicts the performance improvement in terms of (a) $\Delta\mathit{PESQ}$, (b) $\Delta\mathit{STOI}$,  (c) $\Delta\mathit{SIR}^{fws}$, and (d) $\Delta\mathit{CD}$ versus $L$ with respect to the reference microphone signal for the original alternating Kalman filters [\ref{TASLP_ISCLP_altKF_SNR_6}], the modified alternating Kalman filters [\ref{TASLP_ISCLP_altKF_SNR_11}], and the ISCLP Kalman filter [\ref{TASLP_ISCLP_altKF_SNR_16}] at $\mathit{SNR} = 25\,$\si{dB}.
Note that in Fig. \ref{fig:altKFL} and in the following figures presenting performance improvements, 
the resolution of the vertical axes is twice as large as in Fig. \ref{fig:altKFSNR}.
Again, the measures show a high degree of agreement.  
We find that in all measures, the original alternating Kalman filters generally yield less improvement and in addition show a stronger dependency on $L$ as compared to the modified alternating Kalman filters and the ISCLP Kalman filter. 
The improvement for both the modified alternating Kalman filters and the ISCLP Kalman filter saturates at roughly $L = 6$.
The original alternating Kalman filters reach the largest improvement between $L = 8$ and $L = 10$.
In terms of (c) $\Delta\mathit{SIR}^{fws}$ and  (d) $\Delta\mathit{CD}$, however, as opposed to the other two algorithms, its performance  decays again for larger values of $L$  \cite{BraunJune2018}. 
Further simulations showed that for all three algorithms, the dependency on $L$ decreases with decreasing $\mathit{SNR}$ values.
This is expected since at low $\mathit{SNR}$ values, the babble noise component $\v(l)$ becomes pre-dominant, which is temporally uncorrelated, cf. Sec. \ref{sec:signal}, and may therefore not be suppressed by the  LP filter.

Fig. \ref{fig:altKFconv} shows the performance improvement in terms of (a) $\Delta\mathit{PESQ}$, (b) $\Delta\mathit{STOI}$,  (c) $\Delta\mathit{SIR}^{fws}$, and (d) $\Delta\mathit{CD}$ versus time $t$ with respect to the reference microphone signal for the original alternating Kalman filters [\ref{TASLP_ISCLP_altKF_SNR_6}], the modified alternating Kalman filters [\ref{TASLP_ISCLP_altKF_SNR_11}], and the ISCLP Kalman filter [\ref{TASLP_ISCLP_altKF_SNR_16}] with $L = 6$ at $\mathit{SNR} = 10\,$\si{dB}.
Again, the measures largely agree.  
We find that after initialization, all algorithms converge after roughly $4\,\si{s}$. 
The speech source position changes at $8\,\si{s}$, cf. Sec. \ref{sec:caseAwithout}, such that the three algorithms have to re-adapt.
In case of the ISCLP Kalman filter, this does not only require adaptation of $\hat{\w}(l)$, but also of the estimate $\hat{\H}_T(l)$, cf. (\ref{eq:g}), (\ref{eq:B}), and Sec. \ref{sec:targetPSDRETFupdate}.
Note that none of the three algorithms is re-initialized after $t = 8\,\si{s}$, but re-adapt themselves, cf. also Sec. \ref{sec:targetPSDRETFupdate} for the ISCLP Kalman filter. 
However, we find that for all three algorithms, convergence speed after the speech source position change is somewhat reduced  as compared to the initial convergence stage.

\subsubsection{Case B}
\label{sec:rescaseB}

\setlength\fwidth{7.14cm}
\begin{figure}[t]
\centering
\hspace*{-0.218cm}
    \setlength\fheight{12cm} 
    \input{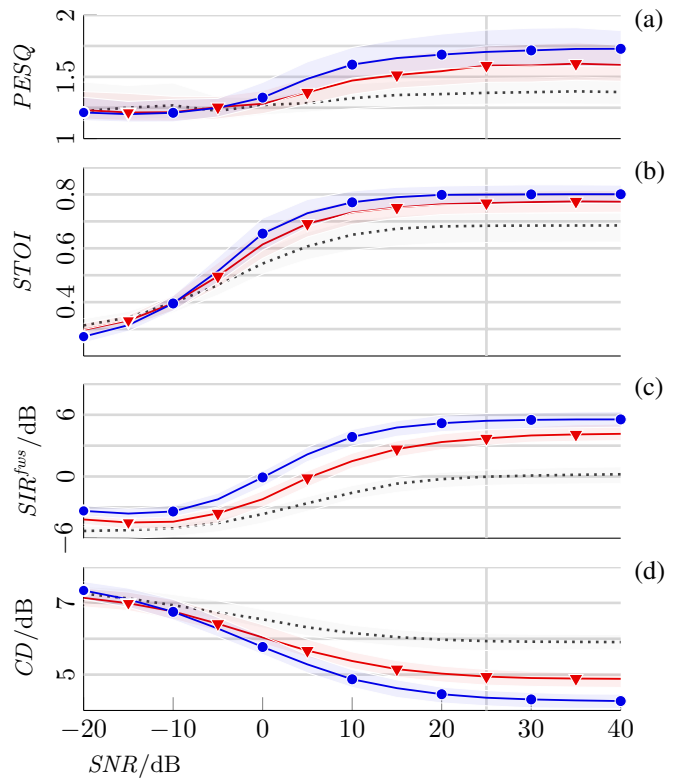} 
\caption{ (a) $\mathit{PESQ}$, (b) $\mathit{STOI}$,  (c) $\mathit{SIR}^{fws}$, and (d) $\mathit{CD}$  versus $\mathit{SNR}$ for the reference microphone signal [\ref{TASLP_ISCLP_altKF_SNR_3}],  the MCLP+GSC Kalman filter cascade [\ref{TASLP_ISCLP_Casc_SNR_8}] and the ISCLP Kalman filter [\ref{TASLP_ISCLP_altKF_SNR_16}] with $L = 6$ if interfering speech is present.
}
\label{fig:cascSNR}
\end{figure}

\setlength\fwidth{7.14cm}
\begin{figure}[t]
\centering
\hspace*{-0.218cm}
    \setlength\fheight{12cm} 
    \input{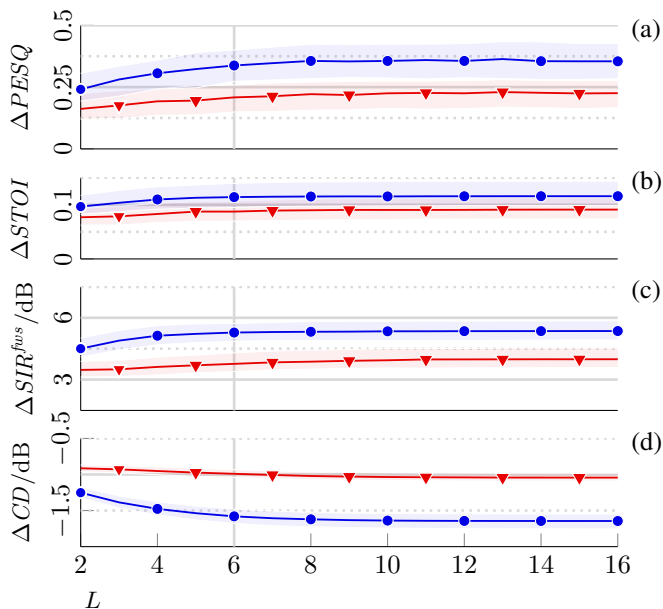} 
\caption{ (a) $\Delta\mathit{PESQ}$, (b) $\Delta\mathit{STOI}$,  (c) $\Delta\mathit{SIR}^{fws}$, and (d) $\Delta\mathit{CD}$ versus $L$ with respect to the reference microphone signal for the MCLP+GSC Kalman filter cascade [\ref{TASLP_ISCLP_Casc_SNR_8}] and the ISCLP Kalman filter [\ref{TASLP_ISCLP_altKF_SNR_16}]  at $\mathit{SNR} = 25\,$\si{dB} if interfering speech is present.
}
\label{fig:cascL}
\end{figure}

Fig. \ref{fig:cascSNR} shows the performance in terms of (a) $\mathit{PESQ}$, (b) $\mathit{STOI}$,  (c) $\mathit{SIR}^{fws}$, and (d) $\mathit{CD}$  versus $\mathit{SNR}$ for the reference microphone signal [\ref{TASLP_ISCLP_altKF_SNR_3}],  the MCLP+GSC Kalman filter cascade [\ref{TASLP_ISCLP_Casc_SNR_8}] and the ISCLP Kalman filter [\ref{TASLP_ISCLP_altKF_SNR_16}] with $L = 6$.
Also here, the measures show a high degree of agreement.  
As in case A, cf. Fig. \ref{fig:altKFSNR}, the reference microphone signal reaches better scores at higher $\mathit{SNR}$ values in all measures.
The curves are, however, generally flatter as compared to those in Fig. \ref{fig:altKFSNR}, which is due to the now additional interfering speech component $\x_2(l)$, cf. Sec. \ref{sec:acoustscen}. 
Above roughly $\mathit{SNR} = -5\,\si{dB}$, both algorithms show a significant improvement over the reference microphone signal in all measures, with the ISCLP Kalman filter clearly outperforming the MCLP+GSC cascade.
For the ISCLP Kalman filter, as compared to case A where $\x_2(l) = \mathbf{0}$, cf. Fig. \ref{fig:altKFSNR}, $\mathit{PESQ}$ now predicts less improvement, while $\mathit{STOI}$ predicts more improvement, indicating different sensitivity of both measures to the additional interfering speech component $\x_2(l)$.

Fig. \ref{fig:cascL} depicts the performance improvement in terms of (a) $\Delta\mathit{PESQ}$, (b) $\Delta\mathit{STOI}$,  (c) $\Delta\mathit{SIR}^{fws}$, and (d) $\Delta\mathit{CD}$ versus $L$ with respect to the reference microphone signal for the MCLP+GSC Kalman filter cascade [\ref{TASLP_ISCLP_Casc_SNR_8}] and the ISCLP Kalman filter [\ref{TASLP_ISCLP_altKF_SNR_16}]  at $\mathit{SNR} = 25\,$\si{dB}.
Again, the ISCLP Kalman filter clearly outperforms the MCLP+GSC Kalman filter cascade in the simulated range. 
For the ISCLP Kalman filter, 
as compared to case A where $\x_2(l) = \mathbf{0}$, cf. Fig. \ref{fig:altKFL}, the improvement shows a stronger dependency on $L$ and saturates somewhat later, indicating that longer filters are required in case of additional temporally correlated components such as $\x_2(l)$, which is in line with the findings in \cite{dietzen19TASLP}.
As in case A, further simulations showed that for both algorithms, the dependency on $L$ decreases with decreasing $\mathit{SNR}$ values.

\section{Conclusion}
\label{sec:conclusion}

In this paper, in order to jointly perform deconvolution and spatial filtering, allowing for dereverberation, interfering speech cancellation and noise reduction, we have presented the ISCLP Kalman filter, which integrates MCLP and the GSC.
Hereat, the SC filter and the LP filter operate in parallel but on different input-data frames, and are estimated jointly.
We further have proposed a spectral Wiener gain post-processor, relating to the Kalman filter's posterior state estimate.
Implementational aspects such as spatio-temporal target component leakage, target PSD estimation and RETF updates, as well as process equation parameter tuning and initialization have been discussed.
The presented ISCLP Kalman filter has been benchmarked in terms of its dependency on the $\mathit{SNR}$ and the filter length $L$, as well as in terms of its convergence behavior.
With $M$ the number of microphones, the ISCLP Kalman filter is roughly $M^2$ times less expensive than both reference algorithms, namely first a pair of alternating Kalman filters in an original and a modified version, and second an MCLP+GSC Kalman filter cascade.
Nonetheless, simulation results indicate better or similar performance as compared to the original or modified version of the former, 
and better performance as compared to the latter.

\section*{Acknowledgments}
\label{sec:acknowledgements}

\footnotesize{This research work was carried out at the ESAT Laboratory of KU Leuven, in the frame of KU Leuven internal funds C2-16-00449, Impulse Fund IMP/14/037; IWT O\&O Project nr. 150611; VLAIO O\&O Project no. HBC.2017.0358; EU FP7-PEOPLE Marie Curie Initial Training Network funded by the European Commission under Grant Agreement no. 316969; the European Union's Horizon 2020 research and innovation program/ERC Consolidator Grant no. 773268. This paper reflects only the authors' views and the Union is not liable for any use that may be made of the contained information.}

\bibliographystyle{IEEEtran}
\bibliography{IEEEabrv}
\end{document}